\pdfoutput=1
\documentclass[12pt,a4paper]{article}
\usepackage{ifthen} % for conditional statements
\newboolean{pdflatex}
\setboolean{pdflatex}{true} % False for eps figures 

\newboolean{articletitles}
\setboolean{articletitles}{true} % False removes titles in references

\newboolean{uprightparticles}
\setboolean{uprightparticles}{false} %True for upright particle symbols

\newboolean{inbibliography}
\setboolean{inbibliography}{false} %True once you enter the bibliography

\usepackage{afterpage}
\usepackage{microtype}
\usepackage{lineno}  % for line numbering during review
\usepackage{xspace} % To avoid problems with missing or double spaces after
\usepackage{caption} %these three command get the figure and table captions automatically small

\usepackage{graphicx}  % to include figures (can also use other packages)
\usepackage{color}
\usepackage{colortbl}
\graphicspath{{./figs/}} % Make Latex search fig subdir for figures

\usepackage{amsmath} % Adds a large collection of math symbols
\usepackage{amssymb}
\usepackage{amsfonts}
\usepackage{upgreek} % Adds in support for greek letters in roman typeset

\newcommand*\patchAmsMathEnvironmentForLineno[1]{%
\expandafter\let\csname old#1\expandafter\endcsname\csname #1\endcsname
\expandafter\let\csname oldend#1\expandafter\endcsname\csname
end#1\endcsname
 \renewenvironment{#1}%
   {\linenomath\csname old#1\endcsname}%
   {\csname oldend#1\endcsname\endlinenomath}%
}
\newcommand*\patchBothAmsMathEnvironmentsForLineno[1]{%
  \patchAmsMathEnvironmentForLineno{#1}%
  \patchAmsMathEnvironmentForLineno{#1*}%
}
\AtBeginDocument{%
\patchBothAmsMathEnvironmentsForLineno{equation}%
\patchBothAmsMathEnvironmentsForLineno{align}%
\patchBothAmsMathEnvironmentsForLineno{flalign}%
\patchBothAmsMathEnvironmentsForLineno{alignat}%
\patchBothAmsMathEnvironmentsForLineno{gather}%
\patchBothAmsMathEnvironmentsForLineno{multline}%
\patchBothAmsMathEnvironmentsForLineno{eqnarray}%
}

\usepackage{hyperref}    % Hyperlinks in references
\usepackage[all]{hypcap} % Internal hyperlinks to floats.

\usepackage{xspace} 
\usepackage{upgreek}

\def\lhcb {\mbox{LHCb}\xspace}

\def\MagUp {\mbox{\em Mag\kern -0.05em Up}\xspace}

\ifthenelse{\boolean{uprightparticles}}%
{

 \def\Pmu         {\ensuremath{\upmu}\xspace}

 \def\Ppi         {\ensuremath{\uppi}\xspace}

 \def\Ppsi        {\ensuremath{\uppsi}\xspace}

 \def\PDelta      {\ensuremath{\Delta}\xspace}                 
 \def\PXi      {\ensuremath{\Xi}\xspace}                 
 \def\PLambda      {\ensuremath{\Lambda}\xspace}                 
 \def\PSigma      {\ensuremath{\Sigma}\xspace}                 
 \def\POmega      {\ensuremath{\Omega}\xspace}                 
 \def\PUpsilon      {\ensuremath{\Upsilon}\xspace}                 
 
 \def\PB      {\ensuremath{\mathrm{B}}\xspace}                 
                  
 \def\PD      {\ensuremath{\mathrm{D}}\xspace}

 \def\PJ      {\ensuremath{\mathrm{J}}\xspace}                 
 \def\PK      {\ensuremath{\mathrm{K}}\xspace}

 \def\Pb      {\ensuremath{\mathrm{b}}\xspace}

 \def\Pi      {\ensuremath{\mathrm{i}}\xspace}

 \def\Pp      {\ensuremath{\mathrm{p}}\xspace}

 \def\Ps      {\ensuremath{\mathrm{s}}\xspace}                 
                  
 \def\Pu      {\ensuremath{\mathrm{u}}\xspace}

}
{

 \def\Pmu         {\ensuremath{\mu}\xspace}

 \def\Ppi         {\ensuremath{\pi}\xspace}

 \def\Ppsi        {\ensuremath{\psi}\xspace}                 
                  
 \mathchardef\PDelta="7101
 \mathchardef\PXi="7104
 \mathchardef\PLambda="7103
 \mathchardef\PSigma="7106
 \mathchardef\POmega="710A
 \mathchardef\PUpsilon="7107
                  
 \def\PB      {\ensuremath{B}\xspace}                 
                  
 \def\PD      {\ensuremath{D}\xspace}

 \def\PJ      {\ensuremath{J}\xspace}                 
 \def\PK      {\ensuremath{K}\xspace}

 \def\Pb      {\ensuremath{b}\xspace}

 \def\Pi      {\ensuremath{i}\xspace}

 \def\Pp      {\ensuremath{p}\xspace}

 \def\Ps      {\ensuremath{s}\xspace}                 
                  
 \def\Pu      {\ensuremath{u}\xspace}

}

\makeatletter
\ifcase \@ptsize \relax% 10pt
  \newcommand{\miniscule}{\@setfontsize\miniscule{4}{5}}% \tiny: 5/6
\or% 11pt
  \newcommand{\miniscule}{\@setfontsize\miniscule{5}{6}}% \tiny: 6/7
\or% 12pt
  \newcommand{\miniscule}{\@setfontsize\miniscule{5}{6}}% \tiny: 6/7
\fi
\makeatother

\DeclareRobustCommand{\optbar}[1]{\shortstack{{\miniscule (\rule[.5ex]{1.25em}{.18mm})}
  \\ [-.7ex] $#1$}}

\def\mup        {{\ensuremath{\Pmu^+}}\xspace}
\def\H      {{\ensuremath{\PH^0}}\xspace}

\def\uquark    {{\ensuremath{\Pu}}\xspace}
\def\uquarkbar {{\ensuremath{\overline \uquark}}\xspace}
\def\uubar     {{\ensuremath{\uquark\uquarkbar}}\xspace}

\def\squark    {{\ensuremath{\Ps}}\xspace}

\def\bquark    {{\ensuremath{\Pb}}\xspace}

\def\pion   {{\ensuremath{\Ppi}}\xspace}

\def\pip    {{\ensuremath{\pion^+}}\xspace}
\def\pim    {{\ensuremath{\pion^-}}\xspace}

\def\kaon    {{\ensuremath{\PK}}\xspace}
  \def\Kbar    {{\kern 0.2em\overline{\kern -0.2em \PK}{}}\xspace}

\def\KorKbar    {\kern 0.18em\optbar{\kern -0.18em K}{}\xspace}

\def\Km      {{\ensuremath{\kaon^-}}\xspace}

  \def\Dbar    {{\kern 0.2em\overline{\kern -0.2em \PD}{}}\xspace}

\def\DorDbar    {\kern 0.18em\optbar{\kern -0.18em D}{}\xspace}

\def\Bbar    {{\ensuremath{\kern 0.18em\overline{\kern -0.18em \PB}{}}}\xspace}

\def\BorBbar    {\kern 0.18em\optbar{\kern -0.18em B}{}\xspace}

\def\Bzb     {{\ensuremath{\Bbar{}^0}}\xspace}

\def\Bsb     {{\ensuremath{\Bbar{}^0_\squark}}\xspace}
\def\Bdb     {{\ensuremath{\Bbar{}^0}}\xspace}

\def\jpsi     {{\ensuremath{{\PJ\mskip -3mu/\mskip -2mu\Ppsi\mskip 2mu}}}\xspace}

  \def\Y#1S{\ensuremath{\PUpsilon{(#1S)}}\xspace}% no space before {...}!

\def\proton      {{\ensuremath{\Pp}}\xspace}

\def\Lz          {{\ensuremath{\PLambda}}\xspace}
\def\Lbar        {{\ensuremath{\kern 0.1em\overline{\kern -0.1em\PLambda}}}\xspace}
\def\LorLbar    {\kern 0.18em\optbar{\kern -0.18em \PLambda}{}\xspace}

\def\Lb      {{\ensuremath{\Lz^0_\bquark}}\xspace}
\def\Lbbar   {{\ensuremath{\Lbar{}^0_\bquark}}\xspace}

\def\to                 {\ensuremath{\rightarrow}\xspace}

\def\CP                {{\ensuremath{C\!P}}\xspace}

\def\AT#1     {\ensuremath{A_{\mathrm{T}}^{#1}}\xspace}           % 2

\def\C#1      {\ensuremath{\mathcal{C}_{#1}}\xspace}                       % 9
\def\Cp#1     {\ensuremath{\mathcal{C}_{#1}^{'}}\xspace}                    % 7
\def\Ceff#1   {\ensuremath{\mathcal{C}_{#1}^{\mathrm{(eff)}}}\xspace}        % 9  
\def\Cpeff#1  {\ensuremath{\mathcal{C}_{#1}^{'\mathrm{(eff)}}}\xspace}       % 7
\def\Ope#1    {\ensuremath{\mathcal{O}_{#1}}\xspace}                       % 2
\def\Opep#1   {\ensuremath{\mathcal{O}_{#1}^{'}}\xspace}                    % 7

\newcommand{\tev}{\ifthenelse{\boolean{inbibliography}}{\ensuremath{~T\kern -0.05em eV}\xspace}{\ensuremath{\mathrm{\,Te\kern -0.1em V}}}\xspace}
\newcommand{\gev}{\ensuremath{\mathrm{\,Ge\kern -0.1em V}}\xspace}
\newcommand{\mev}{\ensuremath{\mathrm{\,Me\kern -0.1em V}}\xspace}
\newcommand{\kev}{\ensuremath{\mathrm{\,ke\kern -0.1em V}}\xspace}
\newcommand{\ev}{\ensuremath{\mathrm{\,e\kern -0.1em V}}\xspace}
\newcommand{\gevc}{\ensuremath{{\mathrm{\,Ge\kern -0.1em V\!/}c}}\xspace}
\newcommand{\mevc}{\ensuremath{{\mathrm{\,Me\kern -0.1em V\!/}c}}\xspace}
\newcommand{\gevcc}{\ensuremath{{\mathrm{\,Ge\kern -0.1em V\!/}c^2}}\xspace}
\newcommand{\gevgevcccc}{\ensuremath{{\mathrm{\,Ge\kern -0.1em V^2\!/}c^4}}\xspace}
\newcommand{\mevcc}{\ensuremath{{\mathrm{\,Me\kern -0.1em V\!/}c^2}}\xspace}
\newcommand{\gevgev}{\ensuremath{{\mathrm{\,Ge\kern -0.1em V^2}}}\xspace}

\def\invfb   {\ensuremath{\mbox{\,fb}^{-1}}\xspace}

\def\gsim{{~\raise.15em\hbox{$>$}\kern-.85em
          \lower.35em\hbox{$\sim$}~}\xspace}
\def\lsim{{~\raise.15em\hbox{$<$}\kern-.85em
          \lower.35em\hbox{$\sim$}~}\xspace}

\def\sPlot{\mbox{\em sPlot}\xspace}
\def\tell1  {TELL1\xspace}
\def\ukl1   {UKL1\xspace}

\newcommand{\eg}{\mbox{\itshape e.g.}\xspace}
\newcommand{\ie}{\mbox{\itshape i.e.}\xspace}

\usepackage{cite} % Allows for ranges in citations
\usepackage{mciteplus}

\usepackage{longtable} % only for template; not usually to be used in PAPERs

\newcommand{\LbJpsippi}{\ensuremath{\Lb\to\jpsi\proton\pim}\xspace}
\def\LambdaStar{{\Lz^*}}
\newcommand{\Pcplus}{\ensuremath{P_c^+}\xspace}
\newcommand{\Zcminus}{\ensuremath{Z_c^-}\xspace}
\def\mppi {\ensuremath{m_{p\pi}}\xspace}
\def\mjpsip  {\ensuremath{m_{\jpsi p}}\xspace}
\def\mjpsipi {\ensuremath{m_{\jpsi \pi}}\xspace}
\def\twolnL {\ensuremath{-2\ln{\cal L}}\xspace}
\def\RpiK{\ensuremath{R_{\pi/K}}\xspace}
\newcommand{\LbJpsipK}{\ensuremath{\Lb\to\jpsi\proton\Km}\xspace}
\newcommand{\Nstar}{\ensuremath{N^{*}}\xspace}

\begin{document}

%%%%%%%%%%%%%%%%%%%%%%%%%
%%%%% Title     %%%%%%%%%
%%%%%%%%%%%%%%%%%%%%%%%%%
\renewcommand{\thefootnote}{\fnsymbol{footnote}}
\setcounter{footnote}{1}

% %%%%%%% CHOOSE TITLE PAGE--------
%\onecolumn
%\input{title-LHCb-INT}
%\input{title-LHCb-ANA}
%\input{title-LHCb-CONF}
\begin{titlepage}
\pagenumbering{roman}

% Header ---------------------------------------------------
\vspace*{-1.5cm}
\centerline{\large EUROPEAN ORGANIZATION FOR NUCLEAR RESEARCH (CERN)}
\vspace*{0.5cm}
\noindent
\begin{tabular*}{\linewidth}{lc@{\extracolsep{\fill}}r@{\extracolsep{0pt}}}
\vspace*{-1.7cm}\mbox{\!\!\!\includegraphics[width=.14\textwidth]{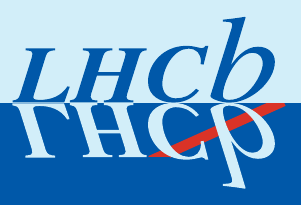}} & & \\
 & & CERN-EP-2016-151 \\  % ID 
 & & LHCb-PAPER-2016-015 \\  % ID 
 & & June 22, 2016 \\ % Date - Can also hardwire e.g.: 23 March 2010
 %& & version 6.0\\
% not in paper \hline
\end{tabular*}

\vspace*{2.0cm}

% Title --------------------------------------------------
{\normalfont\bfseries\boldmath\huge
\begin{center}
  Evidence for exotic hadron contributions to $\Lb \to \jpsi p \pim$ decays 
\end{center}
}

\vspace*{1.0cm}

% Authors -------------------------------------------------
\begin{center}
%In the footnote, replace 'paper' by 'letter' in case of submission to PRL or PLB 
The LHCb collaboration\footnote{Authors are listed at the end of this paper.}
\end{center}

\vspace{\fill}

% Abstract -----------------------------------------------
\begin{abstract}
\noindent
A full amplitude analysis of \LbJpsippi decays is performed 
with a data sample acquired with the \lhcb detector from 7 and 8~TeV $pp$ collisions, 
corresponding to an integrated luminosity of $3\invfb$.
A significantly better description of the data is achieved when, in addition to
the previously observed nucleon excitations $N\to p\pi^-$,
either the $P_c(4380)^+$ and $P_c(4450)^+\to\jpsi p$ states, 
previously observed in $\Lb\to\jpsi p K^-$ decays,  
or the $Z_c(4200)^-\to\jpsi\pi^-$ state, 
previously reported in $B^0\to\jpsi K^+\pi^-$ decays,
or all three, are included in the amplitude models.
The data support a model containing all three exotic states, with a significance of more than three standard deviations. 
%The combined significance of these three exotic states together is more than $3\,\sigma$. 
Within uncertainties, the data are consistent 
with the $P_c(4380)^+$ and $P_c(4450)^+$ production rates expected 
from their previous observation taking account of Cabibbo suppression.
\end{abstract}

\vspace*{1.0cm}

\begin{center}
  Phys.~Rev.~Lett.~{\bf117} (2016) 082003.
\end{center}

\vspace{\fill}

{\footnotesize 
\centerline{\copyright~CERN on behalf of the \lhcb collaboration, licence \href{http://creativecommons.org/licenses/by/4.0/}{CC-BY-4.0}.}}
\vspace*{2mm}

\end{titlepage}

%%%%%%%%%%%%%%%%%%%%%%%%%%%%%%%%
%%%%%  EOD OF TITLE PAGE  %%%%%%
%%%%%%%%%%%%%%%%%%%%%%%%%%%%%%%%

%  empty page follows the title page ----
\newpage
\setcounter{page}{2}
\mbox{~}

\cleardoublepage

%\twocolumn
% %%%%%%%%%%%%% ---------

\renewcommand{\thefootnote}{\arabic{footnote}}
\setcounter{footnote}{0}

%%%%%%%%%%%%%%%%%%%%%%%%%%%%%%%%
%%%%%  Table of Content   %%%%%%
%%%%%%%%%%%%%%%%%%%%%%%%%%%%%%%%
%%%% Uncomment next 2 lines if desired
%\tableofcontents
%\cleardoublepage

%%%%%%%%%%%%%%%%%%%%%%%%%
%%%%% Main text %%%%%%%%%
%%%%%%%%%%%%%%%%%%%%%%%%%

\pagestyle{plain} % restore page numbers for the main text
\setcounter{page}{1}
\pagenumbering{arabic}

%\linenumbers

From the birth of the quark model, it has been anticipated that baryons could be constructed not only from three quarks, but also four quarks and an antiquark \cite{GellMann:1964nj,Zweig:1964}, hereafter referred to as pentaquarks \cite{Montanet:1980te,Lipkin:1987sk}. The distribution of the $\jpsi p$ mass ($m_{\jpsi p}$) in $\Lb\to\jpsi p K^-$, $\jpsi\to\mu^+\mu^-$ decays (charge conjugation is implied throughout the text) observed with the \lhcb detector at the LHC shows a narrow peak suggestive of $uudc\bar c$ pentaquark formation, amidst the dominant formation of various excitations of the $\Lz$ $[uds]$ baryon ($\LambdaStar$) decaying to $K^-p$ \cite{LHCb-PAPER-2015-029,LHCb-PAPER-2015-032}. It was demonstrated that these data cannot be described with $K^-p$ contributions alone without a specific model of them\cite{LHCb-PAPER-2016-009}. Amplitude model fits were also performed on all relevant masses and decay angles of the six-dimensional data \cite{LHCb-PAPER-2015-029}, using the helicity formalism and Breit--Wigner amplitudes to describe all resonances. In addition to the previously well-established $\LambdaStar$ resonances, two pentaquark resonances, named the $P_c(4380)^+$ ($9\,\sigma$ significance) and $P_c(4450)^+$ ($12\,\sigma$), are required in the model for a good description of the data \cite{LHCb-PAPER-2015-029}. The mass, width, and fractional yields (fit fractions) were determined to be $4380\pm8\pm29$ \mev, $205\pm18\pm86$ \mev, $(8.4\pm0.7\pm4.3)\%$, %for $P_c(4380)^+$
and $4450\pm2\pm3$ \mev, $39\pm5\pm19$ \mev, $(4.1\pm0.5\pm1.1)\%$, %for $P_c(4450)^+$,
respectively. Observations of the same two $\Pcplus$ states in another decay would strengthen their interpretation as genuine exotic baryonic states, rather than kinematical effects related to the so-called triangle singularity~\cite{Guo:2015umn,*Liu:2015fea,*Mikhasenko:2015vca}, as pointed out in Ref.~\cite{Burns:2015dwa,*Wang:2015pcn}.% (mass and momentum units in which $c=1$ are used).

\begin{figure}[!tbp]
\centering
\includegraphics[width=0.65\textwidth]{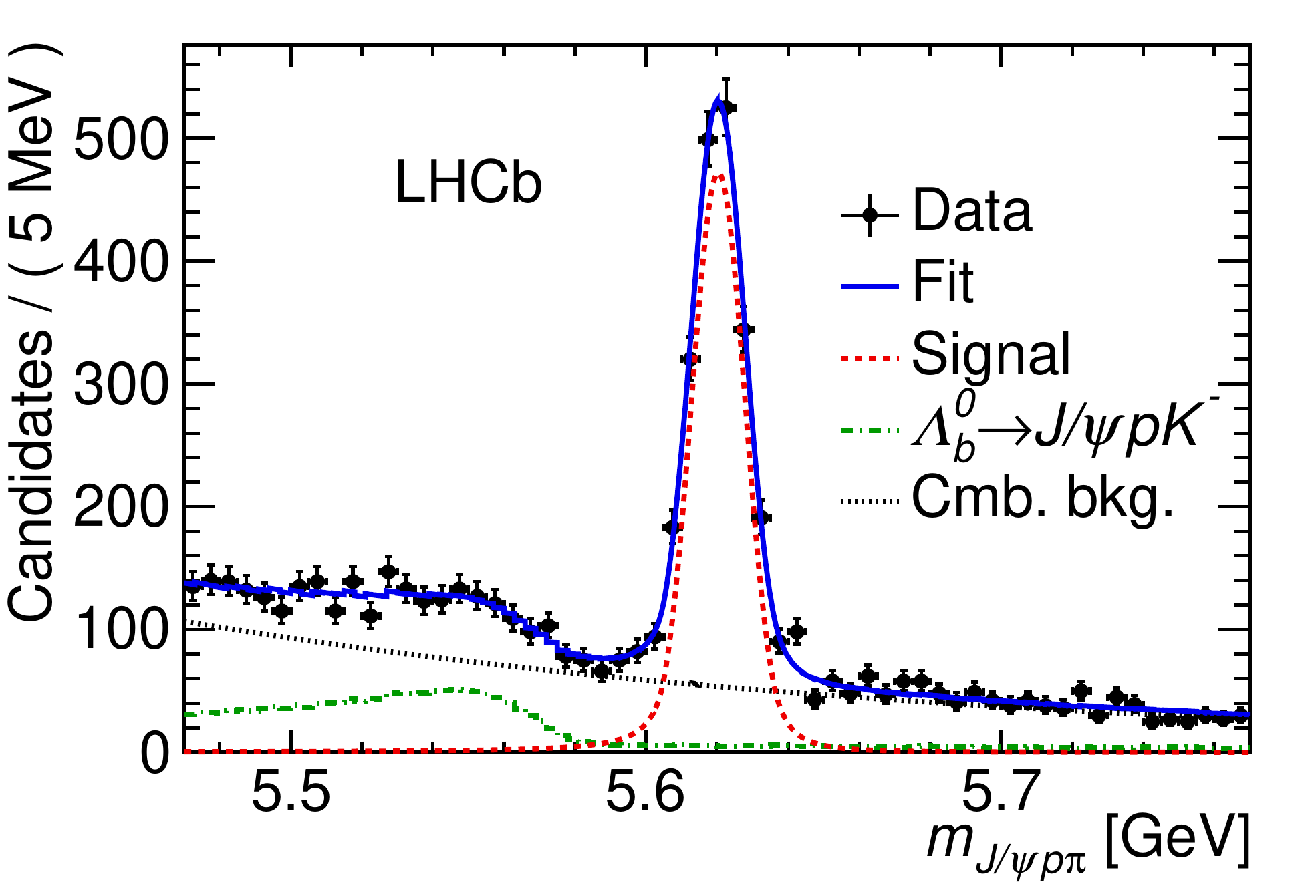}
\caption{Invariant mass spectrum for the selected $\Lb\to \jpsi p \pi^-$ candidates.} 
\label{fig:MassFit}
\end{figure}

In this Letter, $\Lb\to\jpsi p \pi^-$ decays are analyzed, which are related to $\Lb\to\jpsi p K^-$ decays via Cabibbo suppression.
\lhcb has measured the relative branching fraction ${\cal B}(\Lb\to \jpsi p \pi^-)/{\cal B}(\Lb \to \jpsi p K^-)=0.0824\pm0.0024\pm0.0042$~\cite{Aaij:2014zoa} with the same data sample as used here, corresponding to $3\invfb$ of integrated luminosity acquired by the \lhcb experiment in $pp$ collisions at 7 and 8\tev
center-of-mass energy.
The \lhcb detector is a single-arm forward spectrometer covering the pseudorapidity range \mbox{$2<\eta<5$}, described in detail in Refs.~\cite{Alves:2008zz,LHCb-DP-2014-002}. The data selection is similar to that described in Ref.~\cite{LHCb-PAPER-2015-029}, with the $K^-$ replaced by a $\pi^-$ candidate.
%The data preselection is the same, except that the minimum requirement on $\Lb$ candidate flight distance is increased to be 3$\sigma$ significant and that on the alignment of the vector from the primary vertex to the \Lb vertex is tightened. 
In the preselection a larger significance for the \Lb flight distance and a
tighter alignment between the \Lb momentum and the vector from the primary to
the secondary vertex are required.
To remove specific $\Bdb$ and $\Bsb$ backgrounds, candidates are vetoed within a $3\,\sigma$ invariant mass window around the corresponding nominal $B$ mass~\cite{PDG} when interpreted as  $\Bdb \to\jpsi \pip K^-$ or as $\Bsb \to \jpsi K^+ K^-$. In addition, residual long-lived $\Lz\to p \pi^-$ background is excluded if the $p\pi^-$ invariant mass ($\mppi$) lies within $\pm5$\mev of the known $\Lz$ mass~\cite{PDG}. The resulting invariant mass spectrum of $\Lb$ candidates is shown in Fig.~\ref{fig:MassFit}. The signal yield is  $1885\pm50$, determined by an unbinned extended maximum likelihood fit to the mass spectrum. The signal is described by a double-sided Crystal Ball function \cite{Skwarnicki:1986xj}. 
%an empirical function comprising a Gaussian core together with power-law tails on both sides. 
%where the signal radiative tail parameters are fixed to values obtained from simulation. 
The combinatorial background is modeled by an exponential function. The background of $\Lb\to\jpsi p K^-$ events is described by a histogram obtained from simulation, with yield free to vary. 
%This fit is used to assign to the candidates weights depending on $\jpsi p \pi^-$ mass using the \sPlot technique \cite{Pivk:2004ty}.
This fit is used to assign weights to the candidates using the \sPlot technique \cite{Pivk:2004ty}, which allows the signal component to be projected out by weighting each event depending on the $\jpsi p \pi^-$ mass.
Amplitude fits are performed by minimizing a 
six-dimensional unbinned negative log-likelihood, $\twolnL$, 
with the background subtracted using these weights and the efficiency folded into the signal probability density function, as discussed in detail in Ref.~\cite{LHCb-PAPER-2015-029}.
 
Amplitude models for the $\Lb \to \jpsi p \pim$ decays are constructed to
examine the possibility of exotic hadron contributions 
from the $P_c(4380)^+$ and $P_c(4450)^+\to\jpsi p$ states 
and from the $Z_c(4200)^-\to\jpsi \pi^-$ state,
previously reported by the Belle collaboration in 
$B^0\to\jpsi K^+\pi^-$ decays~\cite{Chilikin:2014bkk}
(spin-parity $J^P=1^+$, mass and width of $4196\,_{-29}^{+31}\,_{-13}^{+17}$ \mev
and $370\pm70\,_{-132}^{+\phantom{0}70}$ \mev, respectively).  
By analogy with kaon decays~\cite{Donoghue:1979mu}, $p\pi^-$ contributions from conventional nucleon excitations (denoted as $N^*$) produced with $\Delta I=1/2$ in $\Lb$ decays are expected to dominate over $\PDelta$ excitations with $\Delta I=3/2$, where $I$ is isospin. 
%Conventional nucleon excitations (denoted as $N^*$) decaying to $p \pim$ 
%are expected to be the dominant amplitude contributions. 
%The contribution of $\Delta^*\to p \pim$ resonances is an isospin violating process, and is expected to be suppressed. 
The decay matrix elements for the two interfering decay chains, 
$\Lb \to \jpsi N^*$, $N^* \to p \pim$ 
and $\Lb\to \Pcplus \pim$, $\Pcplus \to \jpsi p$ 
with $\jpsi \to \mup \mu^-$ in both cases, 
are identical to those used in the $\Lb \to \jpsi p K^-$ analysis \cite{LHCb-PAPER-2015-029}, 
with $K^-$ and $\LambdaStar$ replaced by $\pim$ and $N^*$. 
The additional decay chain, $\Lb \to \Zcminus p$, $\Zcminus \to \jpsi \pim$, 
is also included and is discussed in detail in the supplemental material.  
Helicity couplings, describing the dynamics of the decays, are expressed in terms of $LS$ couplings~\cite{LHCb-PAPER-2015-029}, where $L$ is the decay orbital angular momentum, and $S$ is the sum of spins of the decay products. This is a convenient way to incorporate parity conservation in strong decays and to allow for reduction of 
the number of free parameters by excluding high $L$ values for phase-space suppressed decays.   

\begin{table}[t]
\centering
\caption{The $N^*$ resonances used in the different fits. 
Parameters are taken from the PDG \cite{PDG}. 
The number of $LS$ couplings is listed in the columns to the right for the two versions (RM and EM) of the $N^*$ model discussed in the text. 
%A zero entry means the state is excluded from the fit. 
To fix overall phase and magnitude conventions, %which otherwise are arbitrary, 
the $N(1535)$ complex coupling of lowest $LS$ is set to (1,0).}
\label{tab:Lstar}
%\vspace{-0.2cm}
%\ifthenelse{\boolean{wordcount}}%
%{}{
\begin{tabular}{lccccc}
\\[-2.5ex] 
State & $J^P$ & Mass (MeV) & Width (MeV)& RM & EM  \\
\hline \\[-2.5ex] 
NR $p\pi$  &1/2$^-$ & - & - & 4  & 4 \\
$N(1440)$ &1/2$^+$ & 1430 & 350 & 3   & 4 \\
$N(1520)$ &3/2$^-$ & 1515 & 115 &3  & 3\\
$N(1535)$ &1/2$^-$ &1535 & 150 &4    & 4\\
$N(1650)$ &1/2$^-$ &  1655 & 140 & 1   & 4\\
$N(1675)$ &5/2$^-$ & 1675 & 150 & 3   & 5\\
$N(1680)$ &5/2$^+$ &  1685 & 130 &-    & 3\\
$N(1700)$ &3/2$^-$ & 1700& 150 &-    & 3\\
$N(1710)$ &1/2$^+$ &1710 & 100 &- & 4\\
$N(1720)$ &3/2$^+$ & 1720 & 250& 3 & 5\\ 
$N(1875)$ &3/2$^-$ & 1875 & 250 & -  & 3\\
$N(1900)$ &3/2$^+$ & 1900 & 200 &-  & 3 \\
$N(2190)$ &7/2$^-$ & 2190& 500 &-  & 3\\  
%$N(2220)$ &9/2$^+$ & 2250 & 400 &-  & -\\
%$N(2250)$ &9/2$^-$ & 2275 & 500 &-   & -\\
%$N(2600)$ &11/2$^-$ &  2600& 650 & - & - \\ 
$N(2300)$ & 1/2$^+$ & 2300 & 340 & -  & 3 \\
$N(2570)$ & 5/2$^-$ & 2570 & 250 & -  & 3 \\ 
%$\ZP(4380)$& 3/2$^-$& 4  & 4 & 4 & 4\\
%$\ZP(4450)$& 5/2$^+$ & 4  & 4 & 4 & 4\\
\hline
\multicolumn{4}{l}{Free parameters}&  40& 106 \\
%\multicolumn{2}{l}{$\twolnL$} &  $-2164$ & $-2207$ & $-2218$ & $-2230$\\\hline
\end{tabular}
%}
\end{table}

Table~\ref{tab:Lstar} lists the $N^*$ resonances considered in the amplitude model of $p\pi^-$ contributions. There are 15 well-established $N^*$ resonances~\cite{PDG}. The high-mass and high-spin states ($9/2$ and $11/2$) are not included, %since they are unlikely to be produced near the upper kinematic limit of the $\mppi$, since at least three for the $L$ value in the $\Lb$ decays is needed. 
since they require $L\ge3$ in the $\Lb$ decay and therefore are unlikely to be produced near the upper kinematic limit of $\mppi$.
Theoretical models of baryon resonances predict many more high-mass states~\cite{Loring:2001kx}, which have not yet been observed. 
Their absence could arise from decreased couplings of the higher $N^*$ excitations to the simple production and decay channels~\cite{Koniuk:1979vw} and possibly also from 
%possibly because of 
experimental difficulties in identifying broad resonances and insufficient statistics
at high masses in scattering experiments.  The possibility of high-mass, low-spin $N^*$ states is explored by including  
two very significant, but unconfirmed, resonances claimed by the BESIII collaboration 
in $\psi(2S)\to p\bar{p}\pi^0$ decays \cite{Ablikim:2012zk}: 
$1/2^+$ $N(2300)$ and $5/2^-$ $N(2570)$. 
A nonresonant $J^P=1/2^-$ $p\pi^-$ $S$-wave component is also included.
Two models, labeled ``reduced" (RM) and ``extended" (EM), are considered and differ in the number of 
resonances and of $LS$ couplings included in the fit as listed in Table~\ref{tab:Lstar}.
The reduced model, used for the central values of fit fractions, includes 
only the resonances and $L$ couplings that give individually significant contributions. 
The systematic uncertainties and the significances for the exotic states are evaluated with the 
extended model by including all well motivated resonances and the maximal number of 
$LS$ couplings for which the fit is able to converge. %Since it produces the largest deviations from the nominal results it is  
%included among the systematic errors due to the $N^*$ model dependence.
%The RM $N^*$ model has 40 free parameters, the EM model has 106. 

%In the extended model, additional 29 high $LS$ partial wave couplings 
%are fixed to zero based on that higher $LS$ contribution is suppressed by angular momentum barrier. 
All $N^*$ resonances are described by Breit--Wigner functions \cite{LHCb-PAPER-2015-029} 
to model their lineshape and phase variation as a function of $m_{p \pi}$, 
except for the $N(1535)$, which is described by a Flatt{\'e} function~\cite{Flatte:1976xu} 
to account for the threshold of the $n\eta$ channel. 
The mass and width are fixed to the values determined from previous experiments \cite{PDG}.
%and listed in Table~\ref{tab:Lstar}. 
The couplings to the $n\eta$ and $p\pi^-$ channels for the $N(1535)$ state 
are determined by the branching fractions of the two channels~\cite{Anisovich:2011fc}. 
The nonresonant $S$-wave component is described with a function that depends inversely on $m^2_{p\pi}$, as this is found to be preferred by the data.  
An alternative description of the $1/2^-$ $p\pi^-$ contributions, 
including the $N(1535)$ and nonresonant components,  is provided
by a $K$-matrix model obtained from multichannel partial wave analysis by the Bonn--Gatchina group~\cite{Anisovich:2009zy,Anisovich:2011fc} and is used
to estimate systematic uncertainties. 

The limited number of signal events and the large number of free parameters in the amplitude fits 
prevent an open-ended analysis of $\jpsi p$ and $\jpsi\pi^-$ contributions.
Therefore, the data are examined only for the presence of the previously observed $P_c(4380)^+$, $P_c(4450)^+$ states \cite{LHCb-PAPER-2015-029} and the claimed 
$Z_c(4200)^-$ resonance~\cite{Chilikin:2014bkk}. 
In the fits, the mass and width of each exotic state are fixed to the reported central values. 
The $LS$ couplings describing $\Pcplus\to\jpsi p$ decays are also fixed to 
the values obtained from the Cabibbo-favored channel.
This leaves four free parameters per $P_c^+$ state for the $\Lb\to \Pcplus \pi^-$ couplings. 
The nominal fits are performed for the most likely $(3/2^-,5/2^+)$ $J^P$ assignment 
to the $P_c(4380)^+$, $P_c(4450)^+$ states \cite{LHCb-PAPER-2015-029}.     
All couplings for the $1^+$ $Z_c(4200)^-$ contribution are allowed to vary (10 free parameters). 
%The data show a preference for the exotic contributions over the conventional $N^*$ resonances.
 
The fits show a significant improvement when exotic contributions are included. 
%With the EM $N^*$ model the value of $\twolnL$ improves by $5.4^2$, $4.3^2$ and $5.6^2$ when the $P_c(4380)^+$, $P_c(4450)^+$ or $Z_c(4200)^-$ state is added, respectively.
%The improvement is $6.3^2$ when adding the two $P_c^+$ states together 
%and increases to $7.0^2$ when including all three exotic states
%simultaneously. 
%The observed differences in the $\twolnL$ values are used to
%quantify significances of the exotic hadron contributions.
%Statistical simulations show that 
%when a spurious exotic fit component is added to the amplitude model,
%$\Delta(\twolnL)$ follows the $\chi^2$ distribution, 
%with the number of degrees of freedom  
%equal to the number of free parameters in the added component (Wilks' theorem~\cite{Wilks:1938dza}).  
%This is expected since the masses and widths of the exotic resonances are fixed. 
When all three exotic contributions are added to the EM $N^*$-only model, the $\Delta(\twolnL)$ value is 49.0, which corresponds to their combined statistical significance of $3.9\,\sigma$. Including the systematic uncertainties discussed later lowers their significance to $3.1\,\sigma$. The systematic uncertainties are included in subsequent significance figures.
Because of the ambiguity between the $P_c(4380)^+$, $P_c(4450)^+$ and $Z_c(4200)^-$
contributions, no single one of them makes a significant difference to the 
model. Adding either state to a model already containing the other two, or 
the two $P_c^+$ states to a model already containing the $Z_c(4200)^-$ contribution, yields 
significances below $1.7\,\sigma$ ($0.4\,\sigma$ for adding the $Z_c(4200)^-$ after the 
two $P_c^+$ states). If the $Z_c(4200)^-$ contribution is assumed to be negligible, adding
the two $P_c^+$ states to a model without exotics yields a significance of
$3.3\,\sigma$. On the other hand, under the assumption that no $P_c^+$ states are produced, adding the $Z_c(4200)^-$ to a model without exotics yields a significance of $3.2\,\sigma$.  
%Because of the ambiguity between the $P_c(4380)^+$, $P_c(4450)^+$ and $Z_c(4200)^-$ contributions, none of them makes a significant difference when removed from the model. Their individual significances, or the significance of the two $P_c^+$ states together, are all less than $1.7\,\sigma$, and the significance of the $\Z_c(4200)^-$ state is $0.4\,\sigma$. Finally, if the production of the $Z_c(4200)^-$ state is assumed to be negligible, the two $P_c^+$ states taken together are $3.3\,\sigma$ significant.  
The significances are determined using Wilks' theorem~\cite{Wilks:1938dza}, the applicability of which has been verified by simulation.

\begin{figure}[!tbp]
\centering
\includegraphics[width=0.6\textwidth]{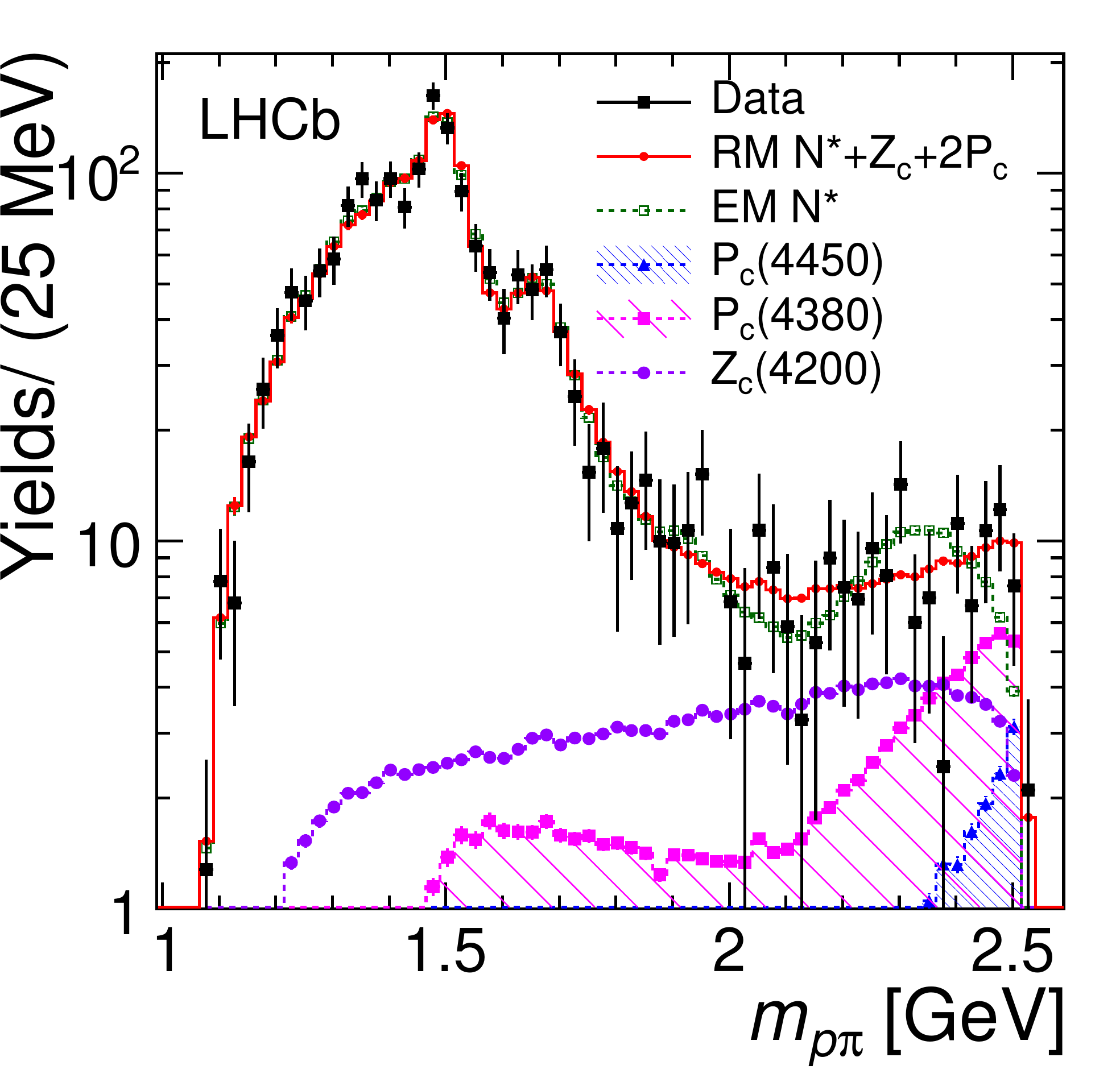}%
\caption{Background-subtracted data and fit projections onto $\mppi$.  Fits are shown with models containing $N^*$ states only (EM) and with $N^*$ states (RM) plus exotic contributions.} 
\label{fig:mppi}
\end{figure}

\begin{figure*}[!tbp]
\centering
\includegraphics[width=0.5\textwidth]{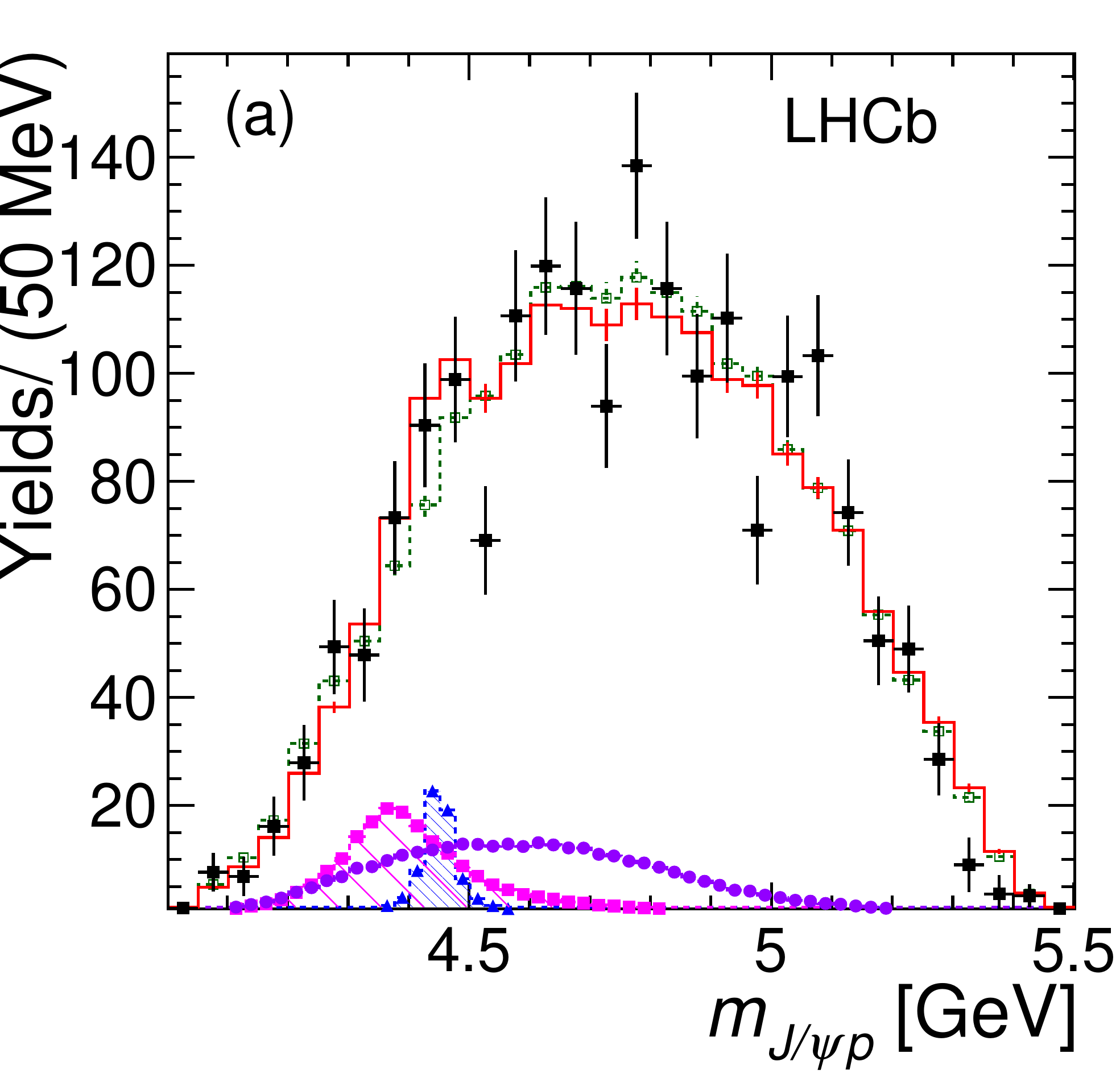}%
\includegraphics[width=0.5\textwidth]{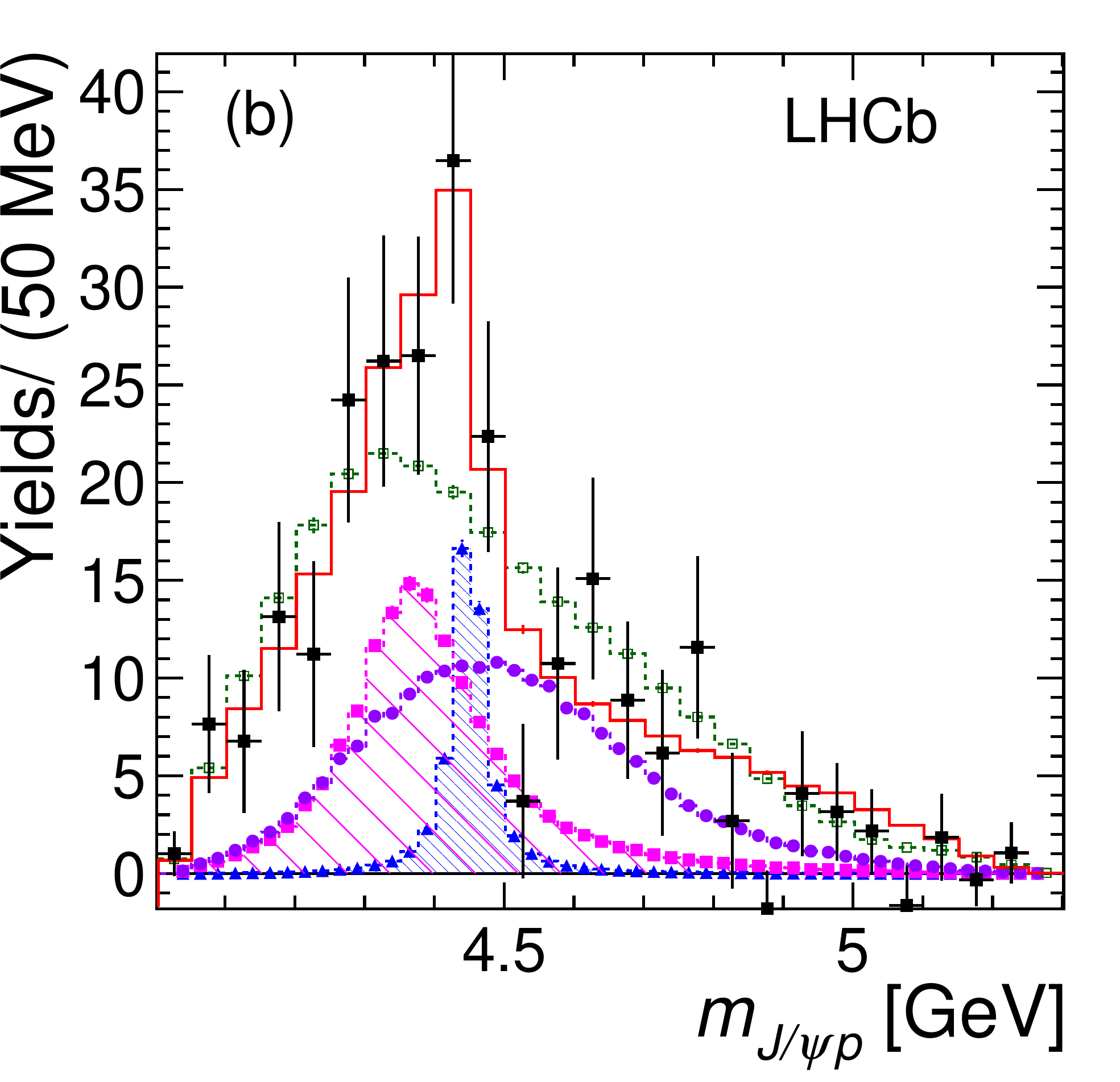}
\caption{Background-subtracted data and fit projections onto $\mjpsip$ for (a) all events and (b) the $\mppi>1.8$ \gev region. 
         See the legend and caption of Fig.~\ref{fig:mppi} for a description of the components.}
\label{fig:mjpsip}
\end{figure*}

\begin{figure*}[!tbp]
\centering
\includegraphics[width=0.5\textwidth]{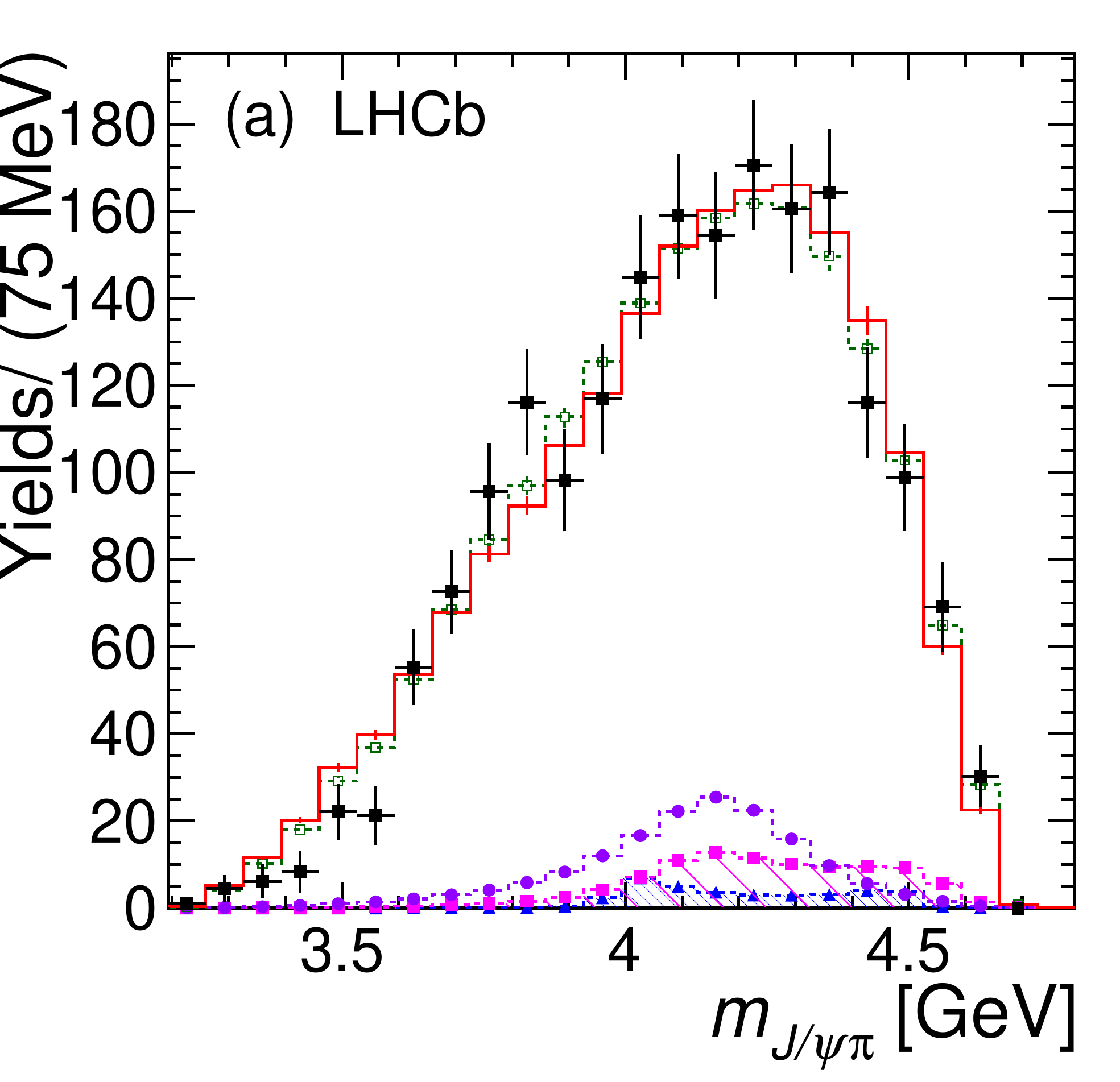}%
\includegraphics[width=0.5\textwidth]{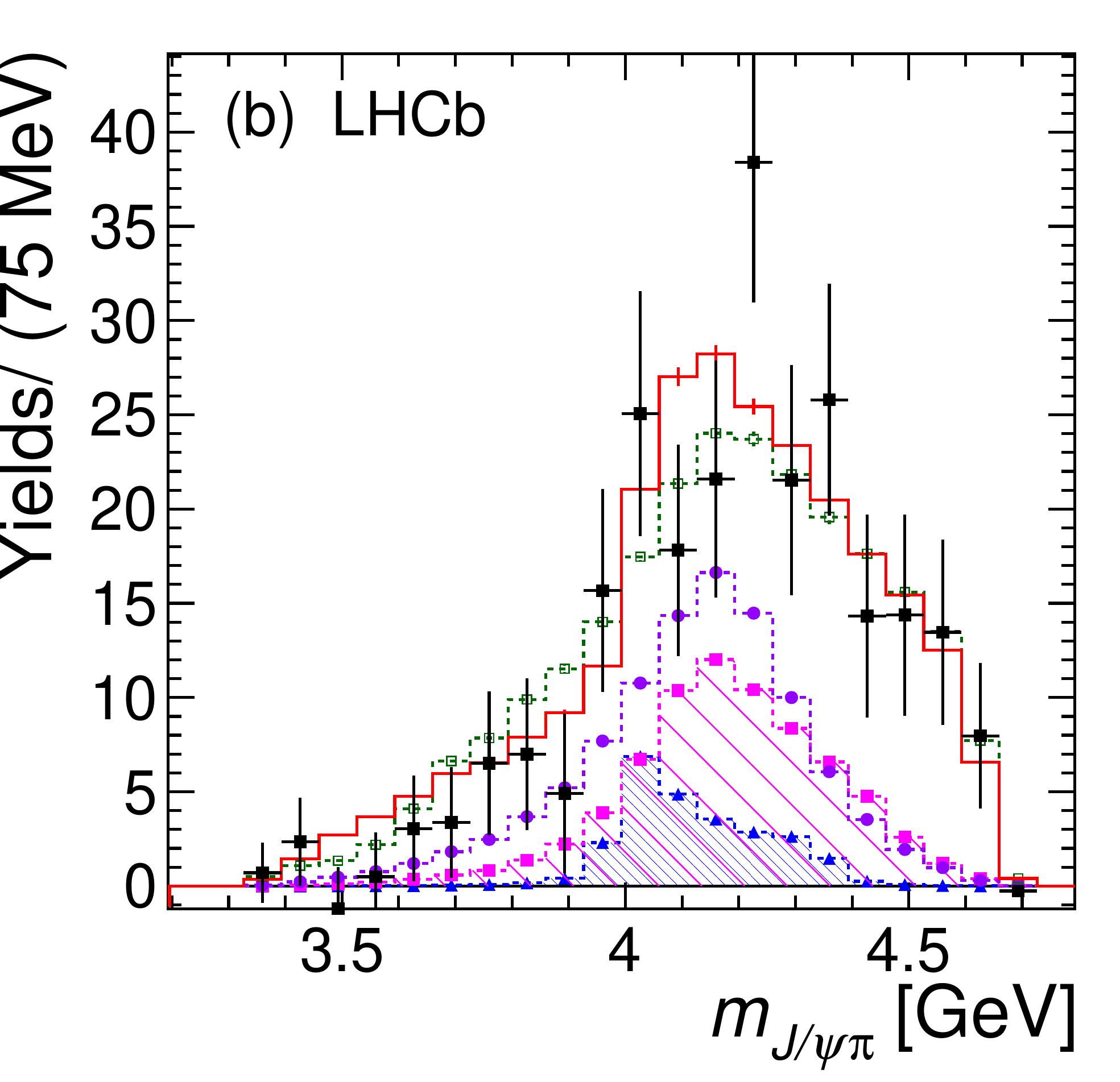}%
\caption{Background-subtracted data and fit projections onto $\mjpsipi$ for (a) all events and (b) the $\mppi>1.8$ \gev region. 
         See the legend and caption of Fig.~\ref{fig:mppi} for a description of the components.}
\label{fig:mjpsipi}
\end{figure*}

\begin{table}[hbt]
\centering
%\vskip 0.3cm
\caption{Summary of absolute systematic uncertainties of the fit fractions in units of percent.}\label{tab:system}
%\vskip -0.2cm
\def\arraystretch{1.2}
%\ifthenelse{\boolean{wordcount}}%
%{}{
\begin{tabular}{lccc}
Source & $P_c(4450)^+$ & $P_c(4380)^+$ & $Z_c(4200)^-$  \\\hline
$N^*$ masses and widths &	$	\pm	0.05 	$&$	\pm	0.23 	$&$	\pm	0.31 	$\\											
$P_c^+$, $Z_c^-$ masses and widths&	$	\pm	0.32 	$&$	\pm	1.27 	$&$	\pm	1.56 	$\\																					
Additional $N^*$&	$_{	-0.23 	}^{+	0.08 	}$&	$_{	-0.55 	}^{+	0.59 	}$&	$_{	-2.92 	}^{+	0.71 	}$\\

Inclusion of $Z_c(4430)^-$ & $+0.01$ & $+0.97$ & $+2.87$\\
Exclusion of $Z_c(4200)^-$ & $-0.15$ & $+1.61$ & - \\
Other $J^P$&	$_{-	0.00 	}^{+	0.38 	}$&	$_{	-0.28 	}^{+	0.92 	}$&	$_{	-2.16 	}^{+	0.00 	}$\\
Blatt--Weisskopf radius &	$	\pm	0.11 	$&$	\pm	0.17 	$&$	\pm	0.21 	$\\											
$L_{\Lb}^{N^*}$ in $\Lb\to\jpsi N^*$&	$	\pm	0.07 	$&$	\pm	0.46 	$&$	\pm	0.04 	$\\											
$L_{\Lb}^{P_c}$ in $\Lb \to P^+_c \pi^-$&	$		-0.05 	$&$		-0.17 	$&$		+0.09 	$\\				
$L_{\Lb}^{Z_c}$ in $\Lb \to Z^-_c p$&	$	\pm	0.07 	$&$	\pm	0.22 	$&$	\pm	0.53 	$\\											
$K$-matrix model	&$		-0.03 	$&$		+0.11 	$&$		-0.02 	$\\											
$P_c^+$ couplings	&$	\pm	0.14 	$&$	\pm	0.31 	$&$	\pm	0.36 	$\\		
Background subtraction & $-0.07$ & $-0.13$ & $-0.39$ \\\hline		
%Total	&$_{	-0.46 	}^{+	0.55 	}$&	$_{	-1.24 	}^{+	1.53 	}$&	$_{	-4.02 	}^{+	1.87 	}$&	$_{	-0.53 	}^{+	0.86 	}$&	$_{	-2.35 	}^{+	1.43 	}$&	$_{	-5.43 	}^{+	3.86 	}$\\
Total&	$_{	-0.48 	}^{+	0.55 	}$&	$_{	-1.58 	}^{+	2.61 	}$&	$_{	-4.04 	}^{+	3.43 	}$\\

\end{tabular}
%}
\end{table}

\begin{figure}[!tbp]
\centering
\includegraphics[width=0.7\textwidth]{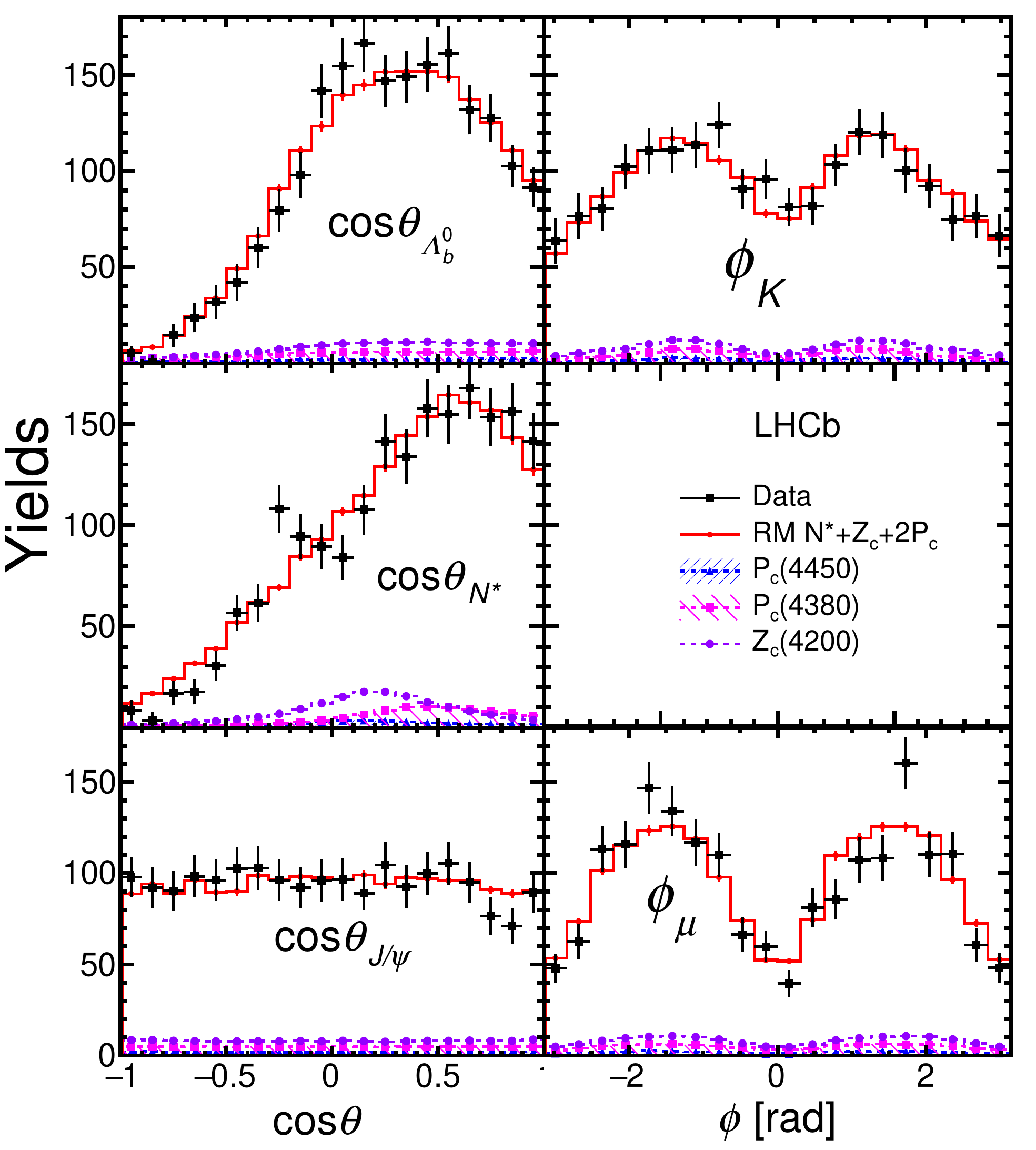}
\caption{Background-subtracted data and fit projections of decay angles describing the $N^*$ decay chain, which are included in the amplitude fit.
The helicity angle of particle $P$, $\theta_P$, is the polar angle in the rest frame of $P$ between a decay product of $P$ and the boost direction from the particle decaying to $P$.
The azimuthal angle between decay planes of $\Lb$ and $N^*$  (of $\jpsi$) is denoted as $\phi_\pi$ ($\phi_\mu$). See Ref.~\cite{LHCb-PAPER-2015-029} for more details.}
\label{fig:angle}
\end{figure}

A satisfactory description of the data is already reached with the RM $N^*$ model 
if either the two $P_c^+$, or the $Z_c^-$, or all three states, are included in the fit.
%This is illustrated in Figs.~\ref{fig:mppi}--\ref{fig:mjpsipi}. 
The projections of the full amplitude fit onto the invariant masses and the decay angles reasonably well reproduce the data, as shown in Figs.~\ref{fig:mppi}--\ref{fig:angle}.
The EM $N^*$-only model does not give good descriptions of the 
peaking structure in $m_{\jpsi p}$ observed for $\mppi>1.8$ \gev (Fig.~\ref{fig:mjpsip}(b)).
In fact, all contributions to $\Delta(\twolnL)$ favoring the exotic components belong to this $\mppi$ region.
The models with the $P_c^+$ states describe the $m_{\jpsi p}$ 
peaking structure better than with the $Z_c(4200)^-$ alone (see the supplemental material).  

The model with all three exotic resonances is used when determining the fit fractions. 
%The reduced $N^*$ model is used for the central values of the fit fractions for exotic 
%resonances, since additional free parameters in the extended $N^*$ models are insignificant.
The sources of systematic uncertainty are listed in Table~\ref{tab:system}.
They include varying the masses and widths of $N^*$  resonances, 
varying the masses and widths of the exotic states, 
considering $N^*$ model dependence and other possible spin-parities $J^P$ for the two $\Pcplus$ states, varying the 
%radius parameter in the Blatt--Weisskopf barrier factor ($d$) 
Blatt--Weisskopf radius~\cite{LHCb-PAPER-2015-029}
between 1.5 and 4.5\,GeV$^{-1}$,  
changing the angular momenta $L$ in $\Lb$ decays that are used in the resonant mass description by one or two units,  
using the $K$-matrix model for the $S$-wave $p\pi$ resonances, varying the fixed couplings of the $\Pcplus$ decay by their uncertainties, and splitting $\Lb$ and $\jpsi$ helicity angles into bins when determining the weights for the background subtraction to account  for correlations between the invariant mass of $\jpsi p \pi^-$ and these angles. 
A putative $Z_c(4430)^-$ contribution~\cite{Chilikin:2013tch,LHCb-PAPER-2014-014,*LHCb-PAPER-2015-038,Chilikin:2014bkk} hardly improves the value of $\twolnL$ relative to the EM $N^*$-only model, and thus is considered among systematic uncertainties. Exclusion of the $Z_c(4200)^-$ state from the fit model is also considered to determine the systematic uncertainties for the two $P_c^+$ states.
%A putative $Z_c(4430)^-$ contribution~\cite{Chilikin:2013tch,LHCb-PAPER-2014-014,*LHCb-PAPER-2015-038,Chilikin:2014bkk} improves the value of $\twolnL$ relative to the EM $N^*$ model by only $7.8$ units with ten additional free parameters, and thus is considered among systematic uncertainties.

The EM model is used to assess the uncertainty due to the $N^*$ modeling when computing significances. The RM model gives larger significances. 
All sources of systematic uncertainties, including the ambiguities in the quantum number assignments to the two $P_c^+$ states, are accounted for in the calculation of the significance of various contributions, by using the smallest $\Delta(\twolnL)$ among the fits representing different systematic variations.
%Since the $J^P$ values of the $P_c^+$ states have not been uniquely determined,
%the fits with the other allowed combinations are also performed and the smallest significance is quoted.
%In addition, uncertainties in masses and widths of the exotic states, slightly reducing the significances, are accounted for. %by taking the smallest significances among the systematic variations.
%In addition, the smallest significance among possible $J^P$ assignments for the two $\Pcplus$ states is used. 
 %are considered by varying them within $\pm\sigma$ of the central values. The maximum reductions are used to give the final significances. 

%Change the width of the exotic states by $-1\sigma$ slightly reduce the significances of these states by at most 7\%, 
%thus doesn't change the conclusion.  
%changing fit region of $m(\jpsi p \pi^-)>5580$\mev to eliminate the $\jpsi K^-p$ reflection background.

The fit fractions for the $P_c(4380)^+$, $P_c(4450)^+$ and $Z_c(4200)^-$ states are measured to be 
$(5.1\pm1.5\,_{-1.6}^{+2.6})\%$, $(1.6\,_{-0.6}^{+0.8}\,_{-0.5}^{+0.6})\%$, and $(7.7\pm2.8\,_{-4.0}^{+3.4})\%$ respectively, and 
to be less than 8.9\%, 2.9\%, and 13.3\% at 90\% confidence level, respectively.
When the two $P_c^+$ states are not considered, 
%which is disfavored by the fit likelihood by 2.3\,$\sigma$, 
the fraction for the $Z_c(4200)^-$ state is surprisingly large, $(17.2\pm3.5)\%$, where the uncertainty is statistical only,
given that its fit fraction was measured to be only $(1.9\,^{+0.7}_{-0.5}\,^{+0.9}_{-0.5})\%$ in $B^0\to\jpsi K^+\pi^-$ decays~\cite{Chilikin:2014bkk}.
%Also noticed is that the $Z_c$ constitutes a large negative interference with the $N^*$ resonances. 
Conversely, the fit fractions of the two $P_c^+$ states remain stable regardless of the inclusion of the $Z_c(4200)^-$ state. 
%Using the results from the default fit and the measured relative branching fraction 
%${\cal B}(\Lb\to \jpsi p \pi^-)/{\cal B}(\Lb \to \jpsi p K^-)=0.0824\pm0.0024\pm0.0042$ from LHCb~\cite{Aaij:2014zoa}, 
We measure the relative branching fraction $\RpiK\equiv {\cal B}(\Lb \to \pi^- P_c^+)/{\cal B}(\Lb \to K^- P_c^+)$ to be
$0.050\pm0.016\,_{-0.016}^{+0.026}\pm0.025$ for $P_c(4380)^+$ and 
$0.033\,_{-0.014}^{+0.016}\,_{-0.010}^{+0.011}\pm0.009$ for $P_c(4450)^+$, respectively, where the first error is statistical, the second is systematic,  
and the third is due to the systematic uncertainty 
on the fit fractions of the $P_c^+$ states in $\jpsi p K^-$ decays. 
The results are consistent with a prediction of (0.07--0.08)~\cite{Cheng:2015cca},
where the assumption is made that an additional diagram with internal $W$ emission, 
which can only contribute to the Cabibbo-suppressed mode, is negligible. 
Our measurement rules out the proposal that the $\Pcplus$ state in the $\LbJpsipK$ decay is produced mainly 
by the charmless $\Lb$ decay via the $b\to\uubar\squark$ transition, since this predicts a very large value for $\RpiK=0.58\pm0.05$~\cite{Hsiao:2015nna}. 

In conclusion, we have performed a full amplitude fit to $\Lb\to \jpsi p \pi^-$ decays
allowing for previously observed conventional ($p\pi^-$) and exotic ($\jpsi p$ and $\jpsi \pi^-$) resonances.
A significantly better description of the data is achieved by either including 
the two $P_c^+$ states observed in $\Lb \to \jpsi p K^-$ decays \cite{LHCb-PAPER-2015-029}, 
or the $Z_c(4200)^-$ state reported by the Belle collaboration in $B^0\to\jpsi\pi^-K^+$ decays~\cite{Chilikin:2014bkk}.
If both types of exotic resonances are included, 
the total significance for them is $3.1\,\sigma$. %Because of the ambiguity between the two types of exotic states, the individual exotic hadron components, or the two $P_c^+$ states taken together, are not significant ($<1.7\,\sigma$).
Individual exotic hadron components, or the two $P_c^+$ states
taken together, are not significant as long as the other(s) is (are) present.
Within the statistical and systematic errors, the data are consistent 
with the $P_c(4380)^+$ and $P_c(4450)^+$ production rates expected 
from their previous observation and Cabibbo suppression. Assuming that the $Z_c(4200)^-$ contribution is negligible, 
there is a $3.3\,\sigma$ significance 
for the two $P_c^+$ states taken together.\hfill \break
 
\noindent 
We thank the Bonn--Gatchina group who provided us with the $K$-matrix $p\pi^-$ model.  
We express our gratitude to our colleagues in the CERN
accelerator departments for the excellent performance of the LHC. We
thank the technical and administrative staff at the LHCb
institutes. We acknowledge support from CERN and from the national
agencies: CAPES, CNPq, FAPERJ and FINEP (Brazil); NSFC (China);
CNRS/IN2P3 (France); BMBF, DFG and MPG (Germany); INFN (Italy); 
FOM and NWO (The Netherlands); MNiSW and NCN (Poland); MEN/IFA (Romania); 
MinES and FANO (Russia); MinECo (Spain); SNSF and SER (Switzerland); 
NASU (Ukraine); STFC (United Kingdom); NSF (USA).
We acknowledge the computing resources that are provided by CERN, IN2P3 (France), KIT and DESY (Germany), INFN (Italy), SURF (The Netherlands), PIC (Spain), GridPP (United Kingdom), RRCKI and Yandex LLC (Russia), CSCS (Switzerland), IFIN-HH (Romania), CBPF (Brazil), PL-GRID (Poland) and OSC (USA). We are indebted to the communities behind the multiple open 
source software packages on which we depend.
Individual groups or members have received support from AvH Foundation (Germany),
EPLANET, Marie Sk\l{}odowska-Curie Actions and ERC (European Union), 
Conseil G\'{e}n\'{e}ral de Haute-Savoie, Labex ENIGMASS and OCEVU, 
R\'{e}gion Auvergne (France), RFBR and Yandex LLC (Russia), GVA, XuntaGal and GENCAT (Spain), Herchel Smith Fund, The Royal Society, Royal Commission for the Exhibition of 1851 and the Leverhulme Trust (United Kingdom).

%\input{appendix}

% This should be taken out in the final paper

\ifx\mcitethebibliography\mciteundefinedmacro
\PackageError{LHCb.bst}{mciteplus.sty has not been loaded}
{This bibstyle requires the use of the mciteplus package.}\fi
\providecommand{\href}[2]{#2}

%\addcontentsline{toc}{section}{References}
%\setboolean{inbibliography}{true}
%\bibliographystyle{LHCb}
%\bibliography{main,LHCb-PAPER,LHCb-CONF,LHCb-DP,LHCb-TDR,Pentaquark}

\newpage

\clearpage
\def\Mat{\mathcal{M}}
\def\ZP{P_c}
\def\ZC{Z_c}
\section*{Appendix: Supplemental material}
\tableofcontents
\newpage
\label{sec:Supplementary-App}
\section{Dalitz plot distributions}
\begin{figure}[!b]
\centering
\includegraphics[width=0.5\textwidth]{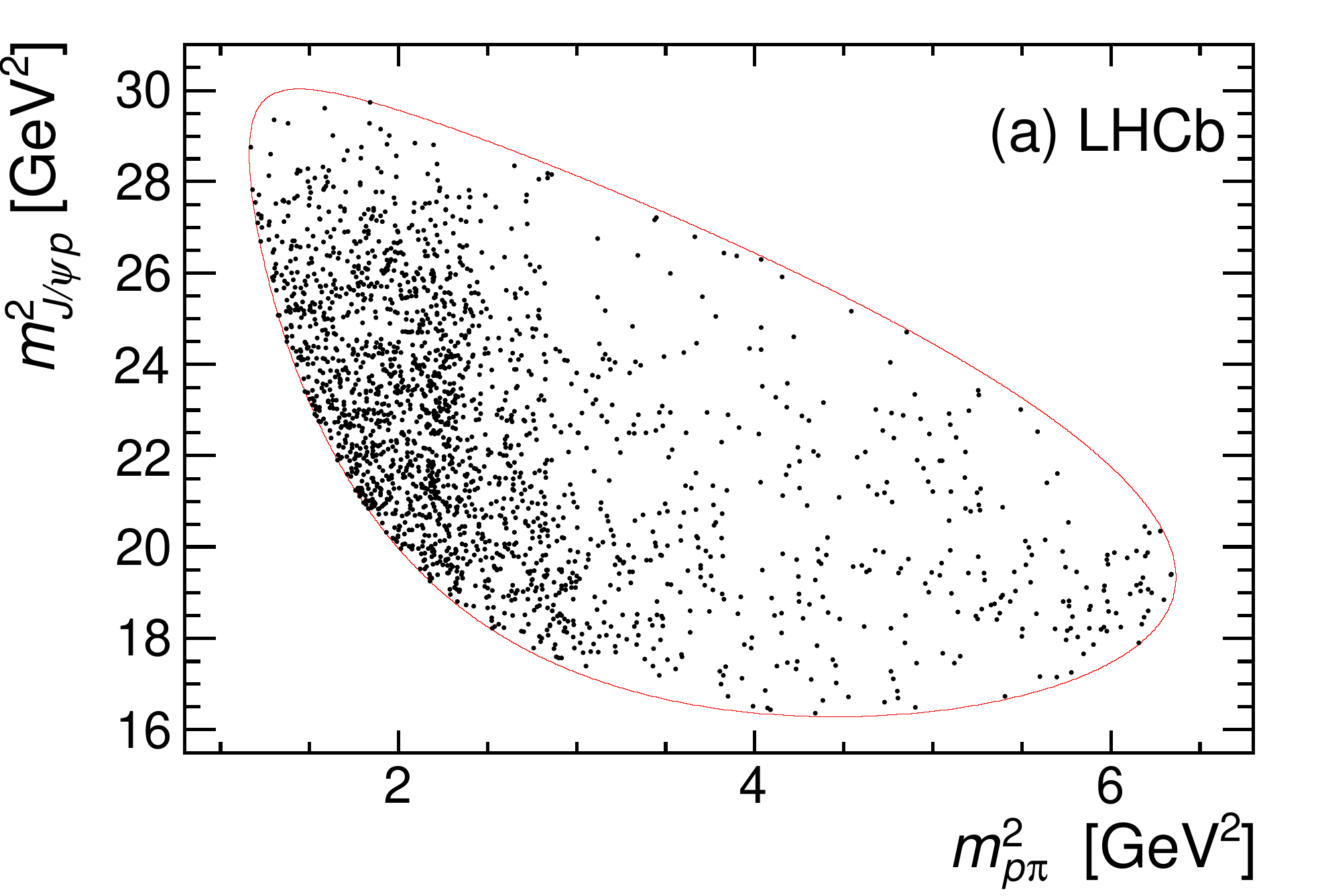}%
\includegraphics[width=0.5\textwidth]{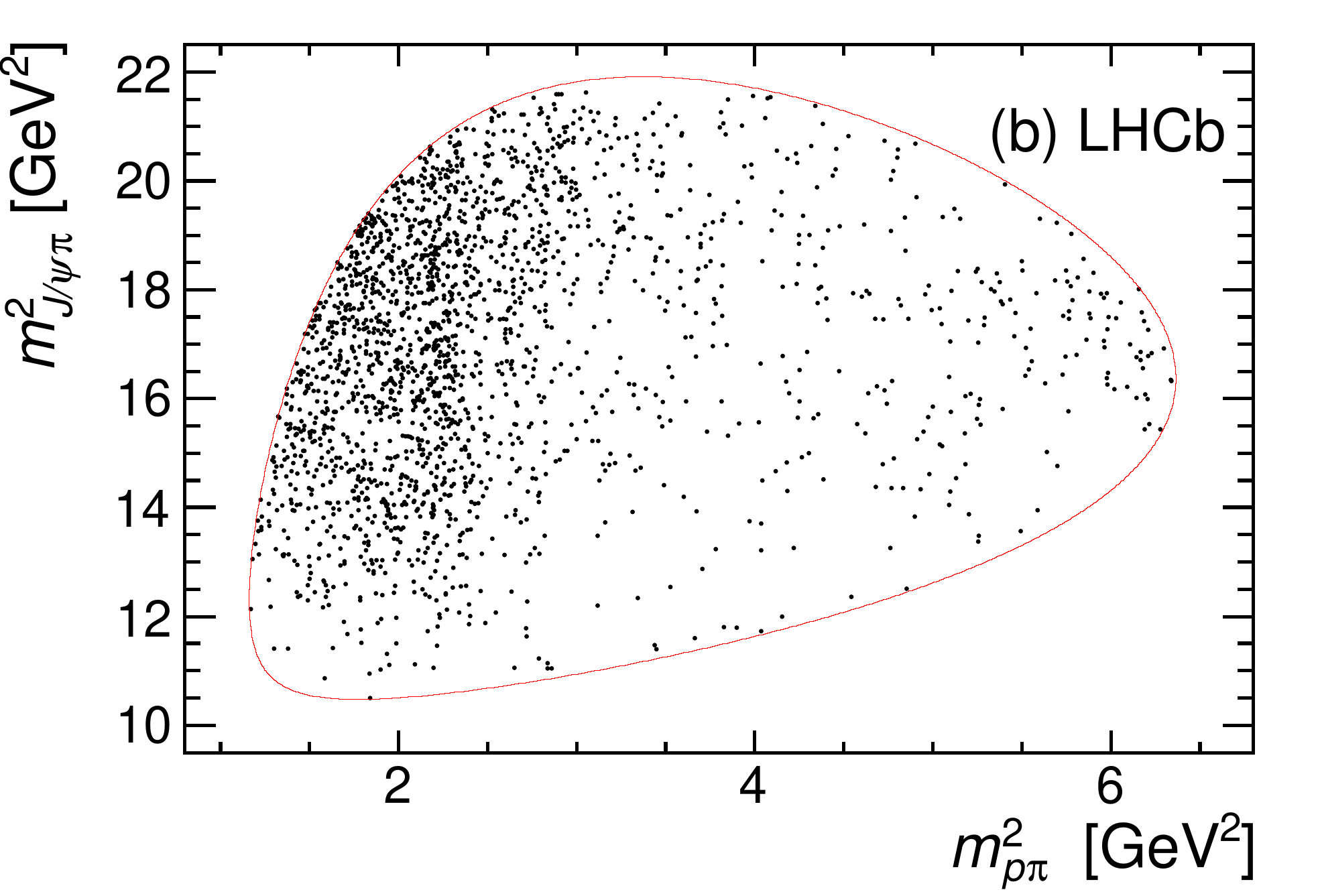}
\caption{Invariant mass squared of $p\pim$ versus either (a) $\jpsi p$ or (b) $\jpsi \pi$ for candidates within $\pm15\mev$ of the $\Lb$ mass, which contain 17\% background. The lines show the kinematic boundaries with the $\Lb$ mass constrained to the known value.}
\label{dlz-point}
\end{figure}

\begin{figure}[!b]
\centering
\includegraphics[width=0.5\textwidth]{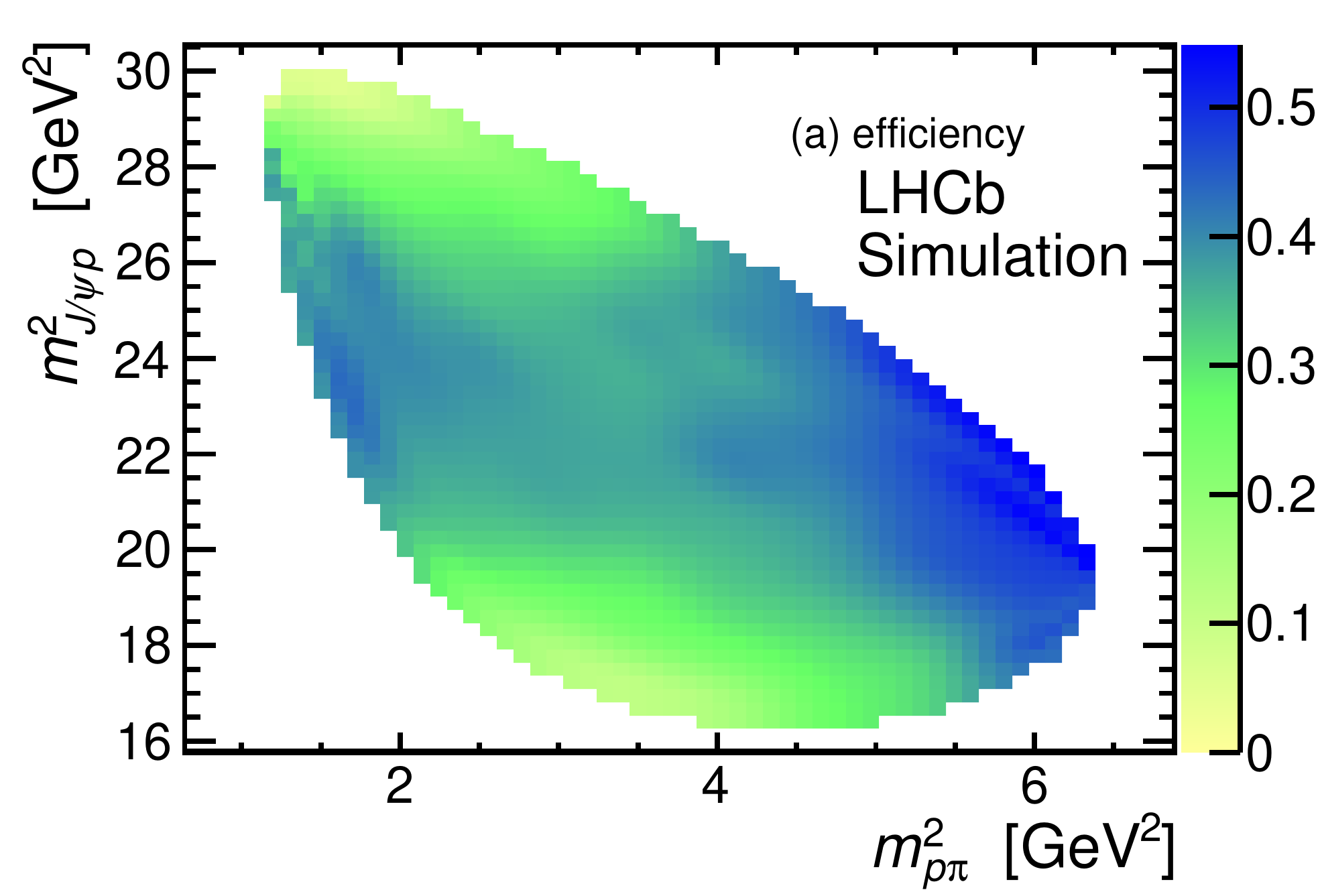}%
\includegraphics[width=0.5\textwidth]{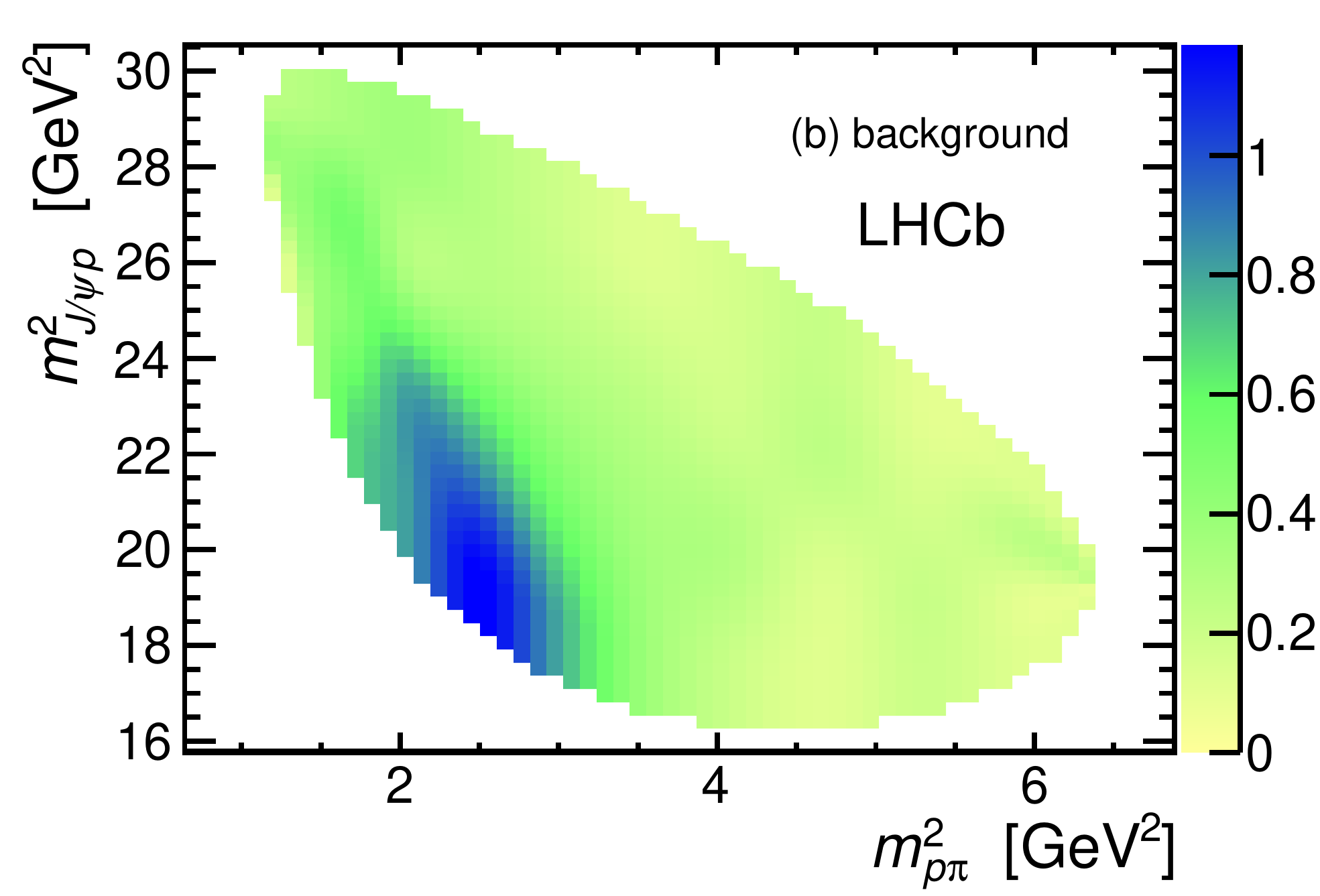}
\caption{(a) Relative signal efficiency and (b) background distribution on the Dalitz plane in arbitrary units.}\label{dlzbg}
\end{figure}

In Fig.~\ref{dlz-point}, we show the Dalitz plots using the invariant mass squared, $m^2_{p\pi}$ vs $m^2_{\jpsi\proton}$,  and $m^2_{p\pi}$ vs $m^2_{\jpsi\pi}$ as independent variables. As expected, significant $\Nstar$ contributions are seen, especially in the region around $m^2_{p\pi}\approx 2\gevgev$. 
There is no visible narrow band around the $P_c(4450)$ region at about $20\gevgev$ in the $\jpsi p$ mass-squared. However many events concentrate in a limited window around $m^2_{p\pi} = 6\gevgev$ and $m^2_{\jpsi p}$ between $18$ to $20\gevgev$, which could be due to the $P_c$ contribution.  Distributions of efficiency and background on the Dalitz plane are shown in Fig.~\ref{dlzbg}, where the background consists of events from the \Lb candidate mass sideband of $5665-5770$\mev. The efficiency is obtained from MC simulation with data-driven corrections applied for the particle identification of the pion and proton.% and for the $\Lb$ kinematics using $\Lb\to\jpsi p K^-$ sample. 

\section{Additional fit results}
%\subsection{Angular projections from reduced model with the three exotic states included}
%In Fig.~\ref{angle} we show the result of the reduced $N^*$ model together with the $\Pcplus$ and $\Zcminus$ contributions onto the angular variables. The data is well described by the fits. 
%\begin{figure}[b]
%\centering
%\includegraphics[width=0.8\textwidth]{angle}
%\caption{Distributions of various decay angles.}
%\label{angle}
%\end{figure}

\subsection{Additional fit displays}
\begin{figure}[!tbp]
\centering
\includegraphics[width=0.6\textwidth]{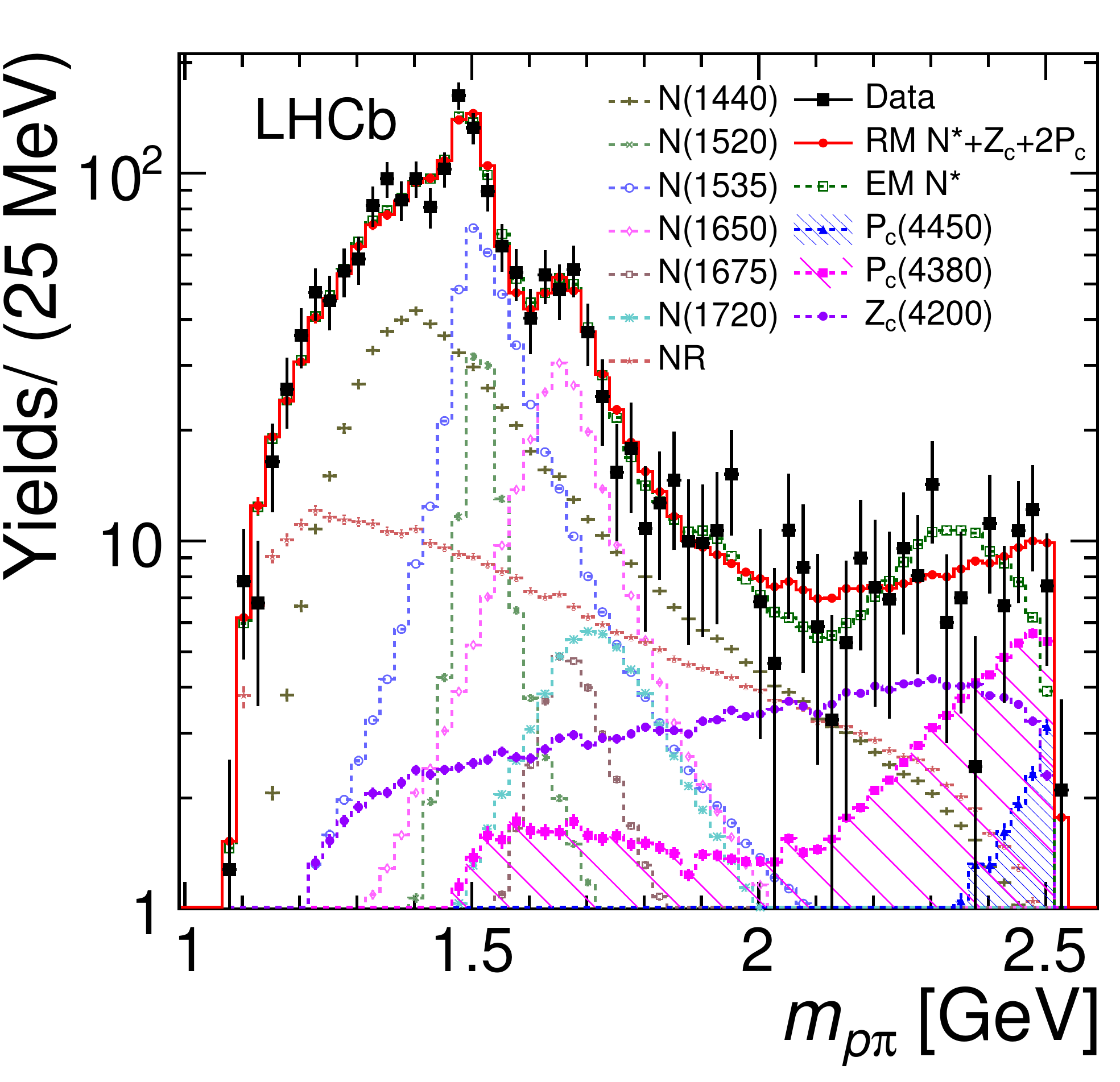}%
\caption{Background-subtracted data and fit projections onto $\mppi$.  Fits are shown with models containing $N^*$ states only (EM) and with $N^*$ states (RM) plus exotic contributions. Individual fit components are shown only for the fit which includes all three exotic resonances. } 
\label{fig:mppi-all}
\end{figure}

\begin{figure}[!tbp]
\centering
\includegraphics[width=0.6\textwidth]{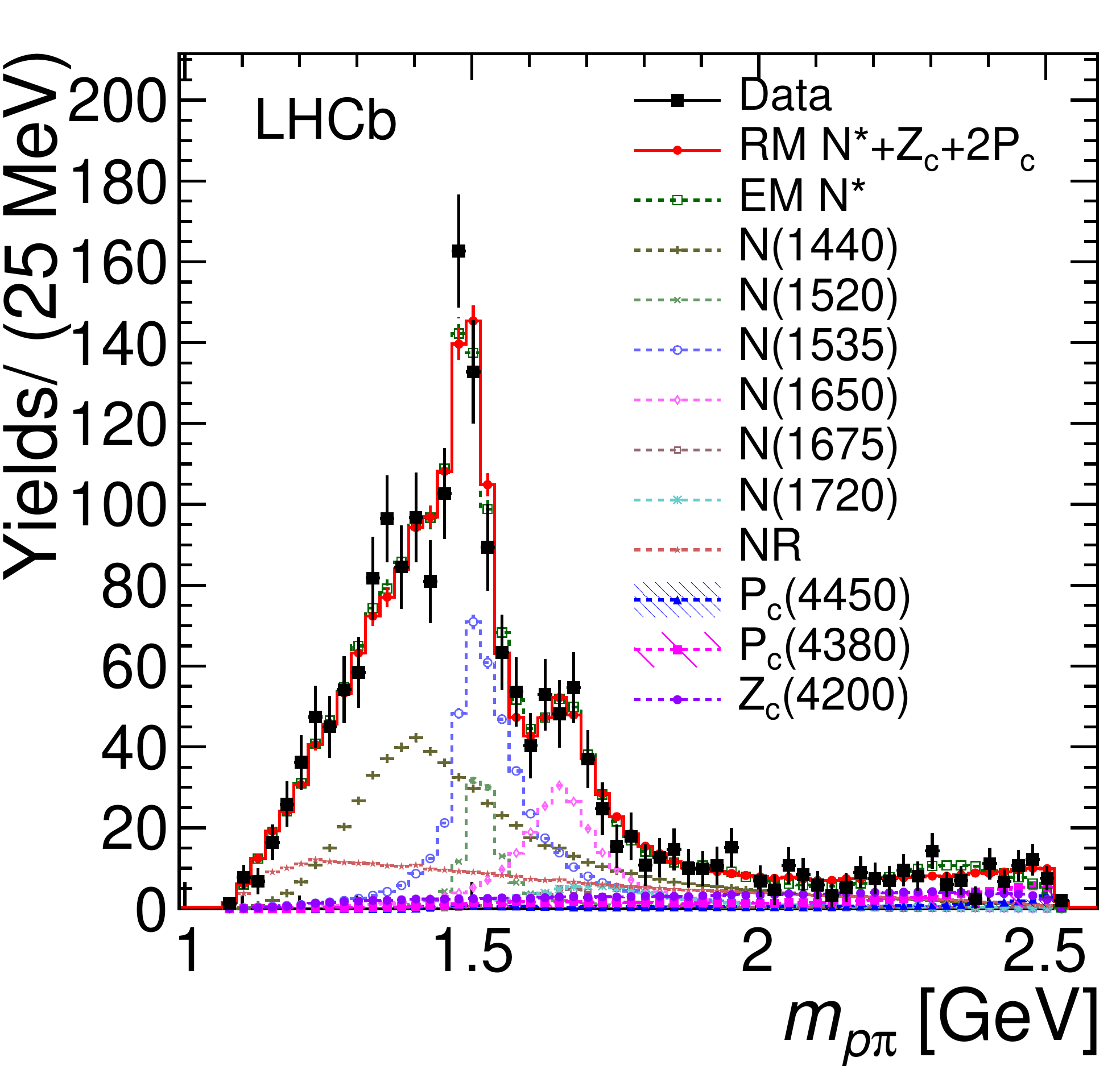}%
\caption{Background-subtracted data and fit projections onto $\mppi$.  Fits are shown with models containing $N^*$ states only (EM) and with $N^*$ states (RM) plus exotic contributions. Individual fit components are shown only for the fit which includes all three exotic resonances. } 
\label{fig:mppi-linear}
\end{figure}

\begin{figure}[!tbp]
\centering
\includegraphics[width=0.6\textwidth]{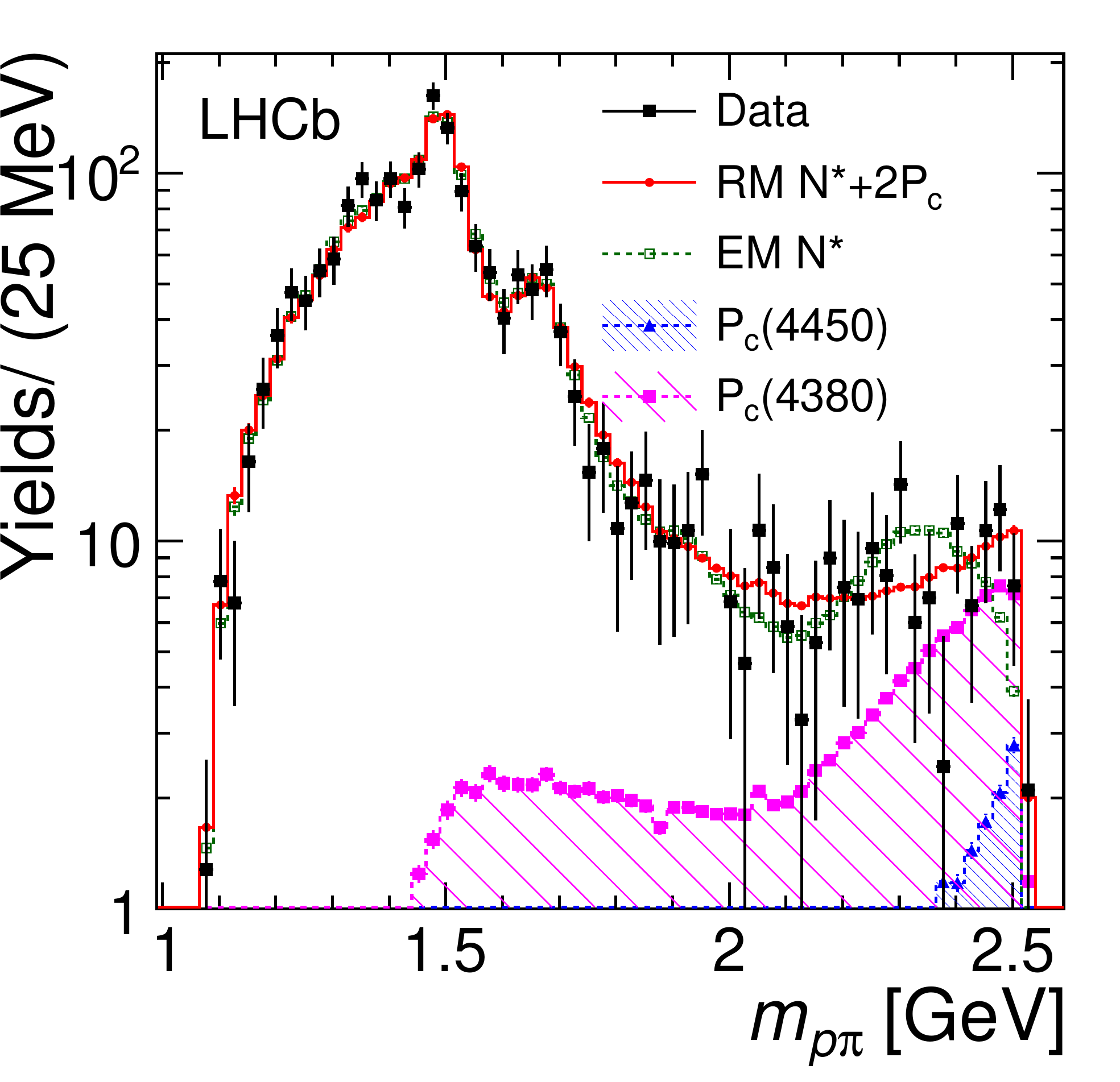}%
\caption{Background-subtracted data and fit projections onto $\mppi$.  Fits are shown with models containing $N^*$ states only (EM) and with $N^*$ states (RM) plus the two $P_c^+$ resonances. } 
\label{fig:mppi-2pc}
\end{figure}

\begin{figure*}[!tbp]
\centering
\includegraphics[width=0.5\textwidth]{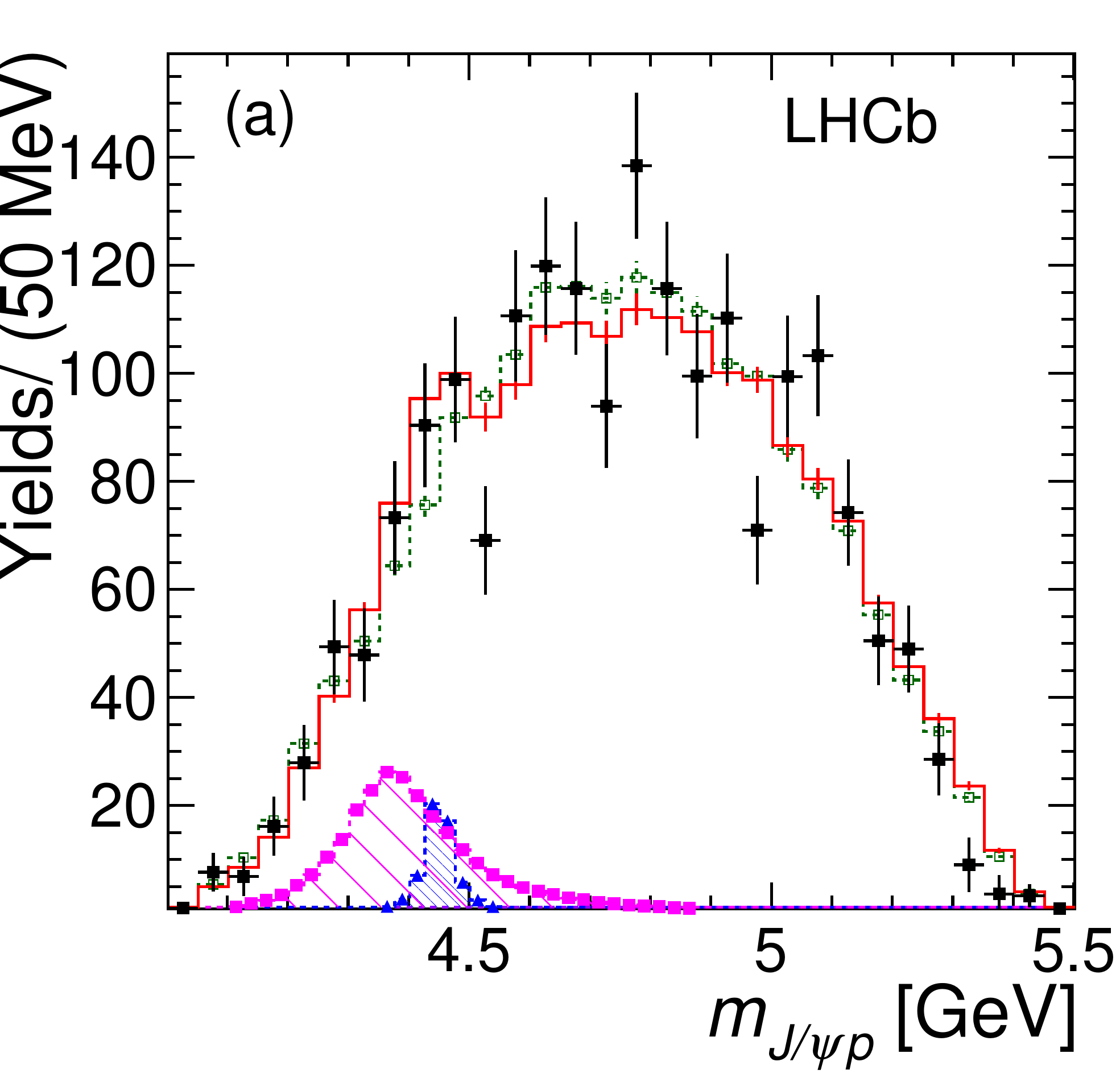}%
\includegraphics[width=0.5\textwidth]{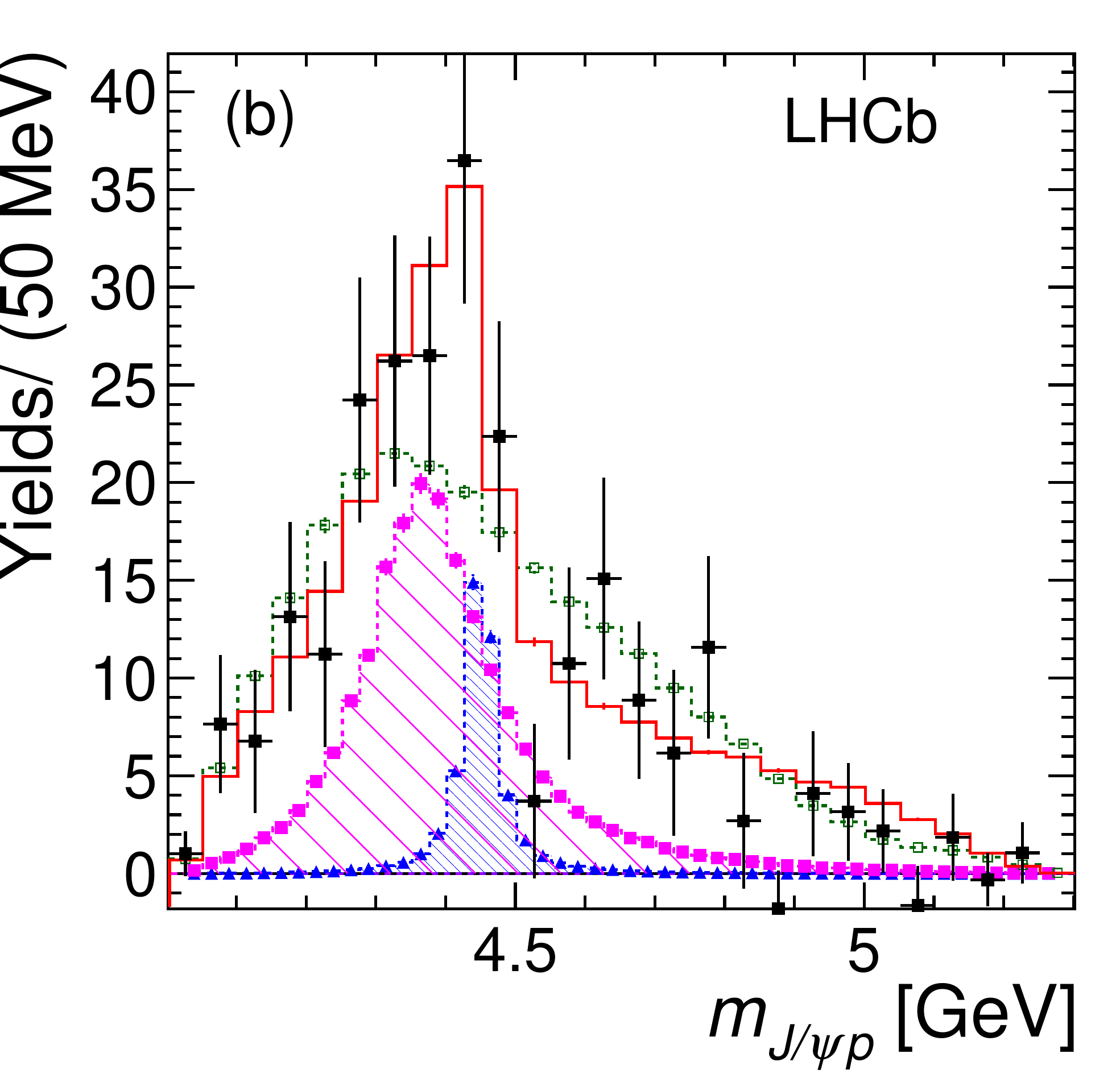}
\caption{Background-subtracted data and fit projections onto $\mjpsip$ for (a) all events and (b) the $\mppi>1.8$ \gev region. 
         See the legend and caption of Fig.~\ref{fig:mppi-2pc} for a description of the components.}
\label{fig:mjpsip-2pc}
\end{figure*}

\begin{figure*}[!tbp]
\centering
\includegraphics[width=0.5\textwidth]{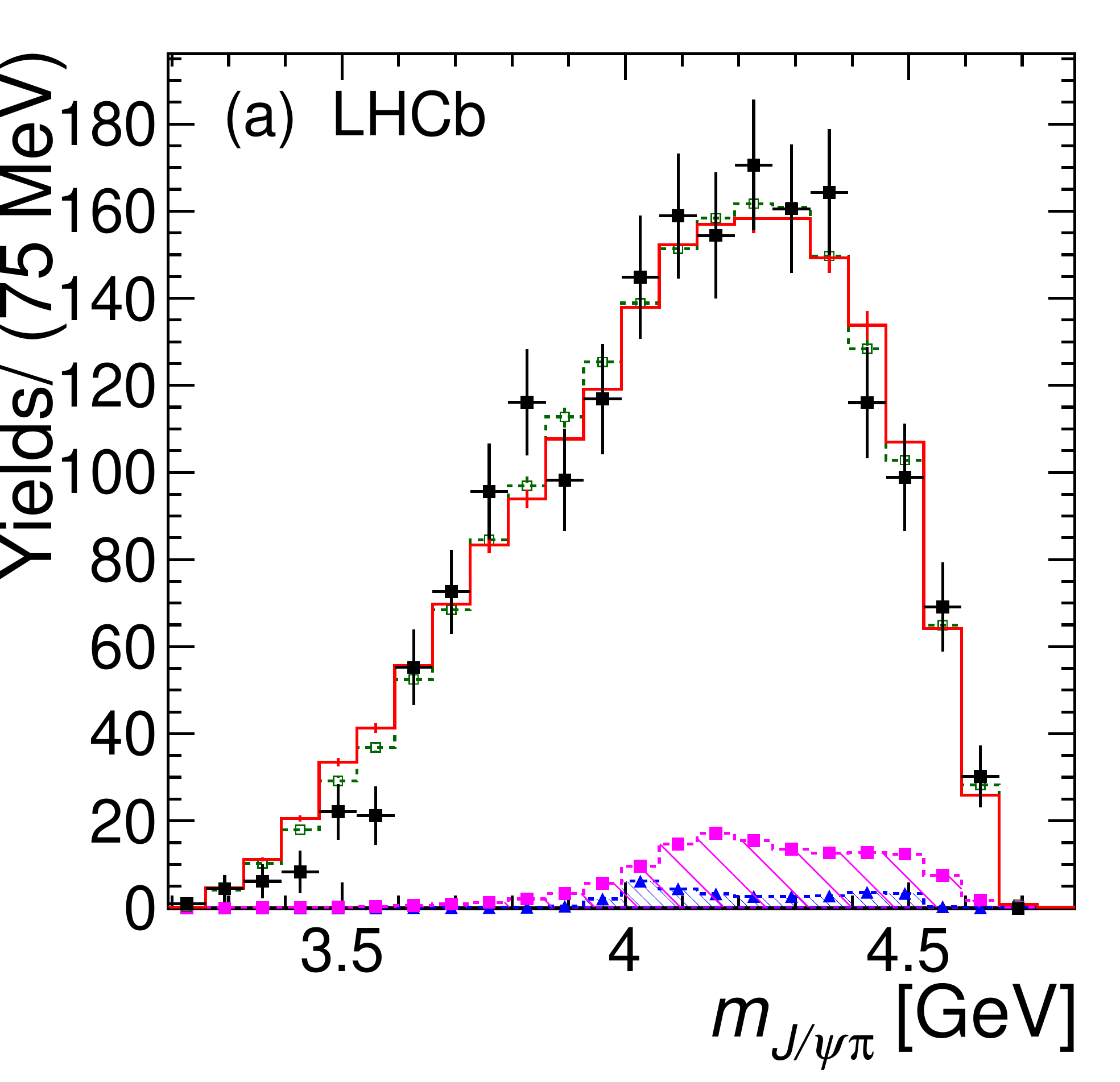}%
\includegraphics[width=0.5\textwidth]{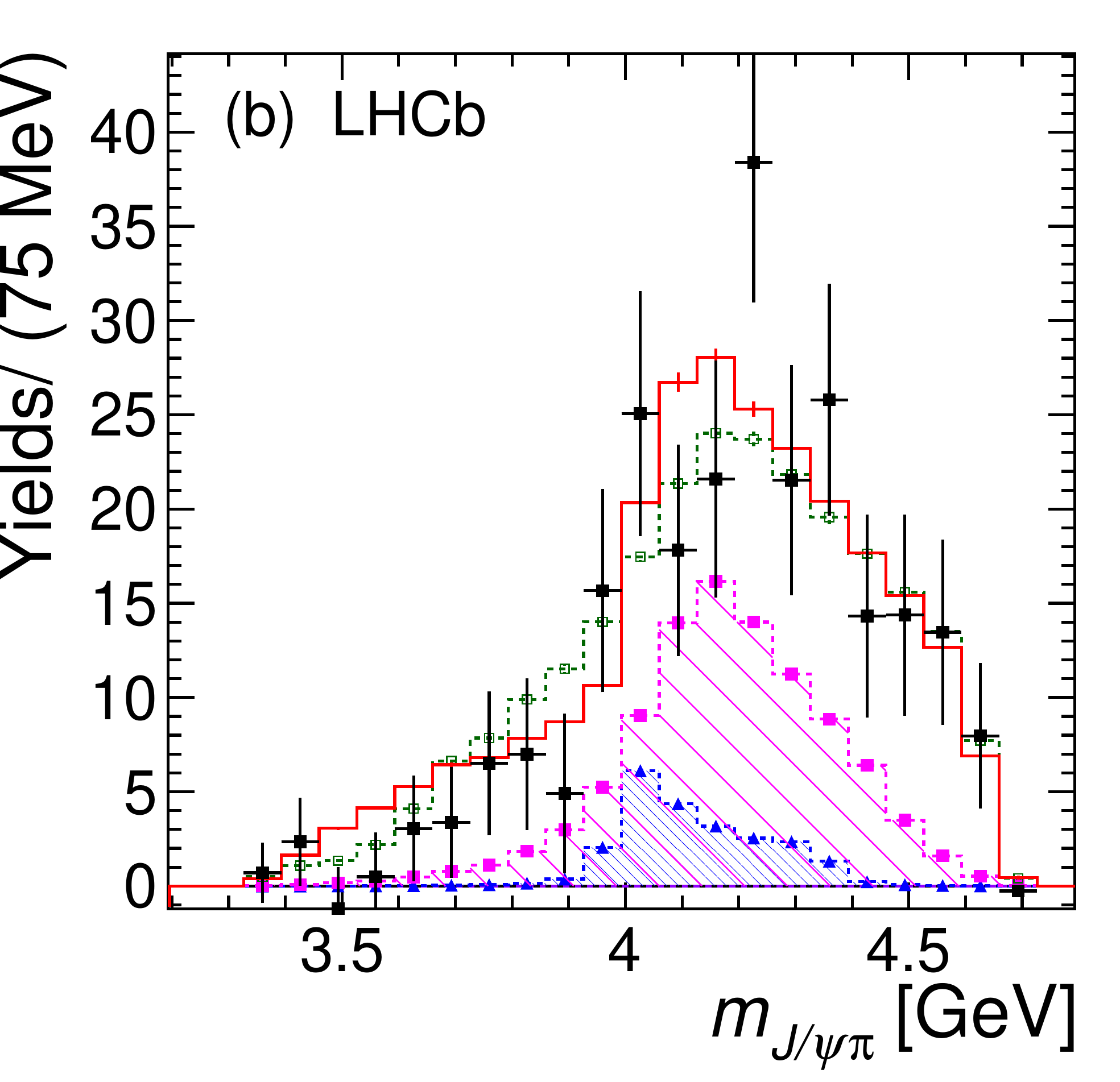}%
\caption{Background-subtracted data and fit projections onto $\mjpsipi$ for (a) all events and (b) the $\mppi>1.8$ \gev region. 
         See the legend and caption of Fig.~\ref{fig:mppi-2pc} for the description.}
\label{fig:mjpsipi-2pc}
\end{figure*}

\begin{figure}[!tbp]
\centering
\includegraphics[width=0.6\textwidth]{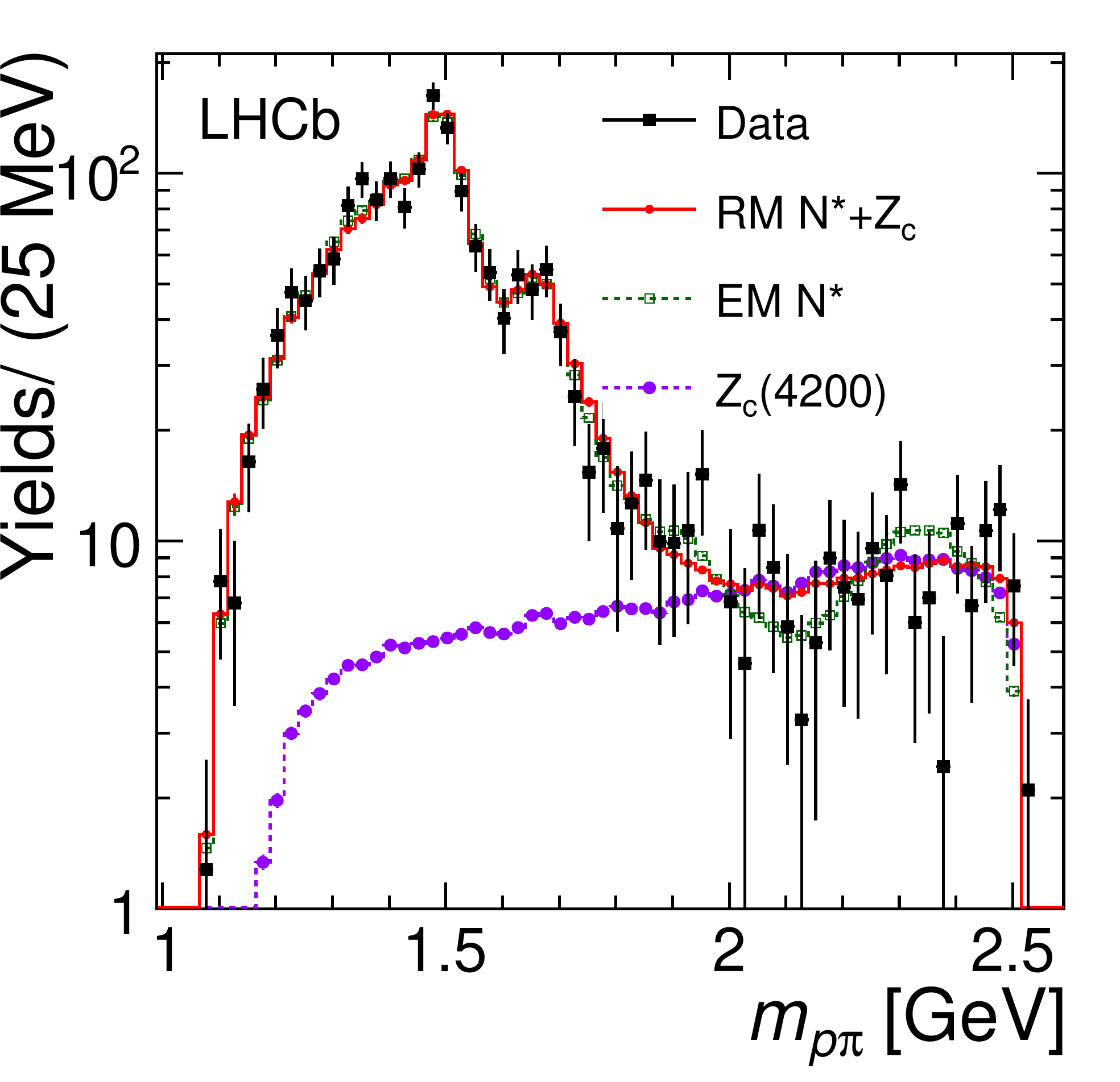}%
\caption{
Background-subtracted data and fit projections onto $\mppi$.  Fits are shown with models containing $N^*$ states only (EM) and with $N^*$ states (RM) plus the $Z_c(4200)^-$ resonance. } 
\label{fig:mppi-zc}
\end{figure}

\begin{figure*}[!tbp]
\centering
\includegraphics[width=0.5\textwidth]{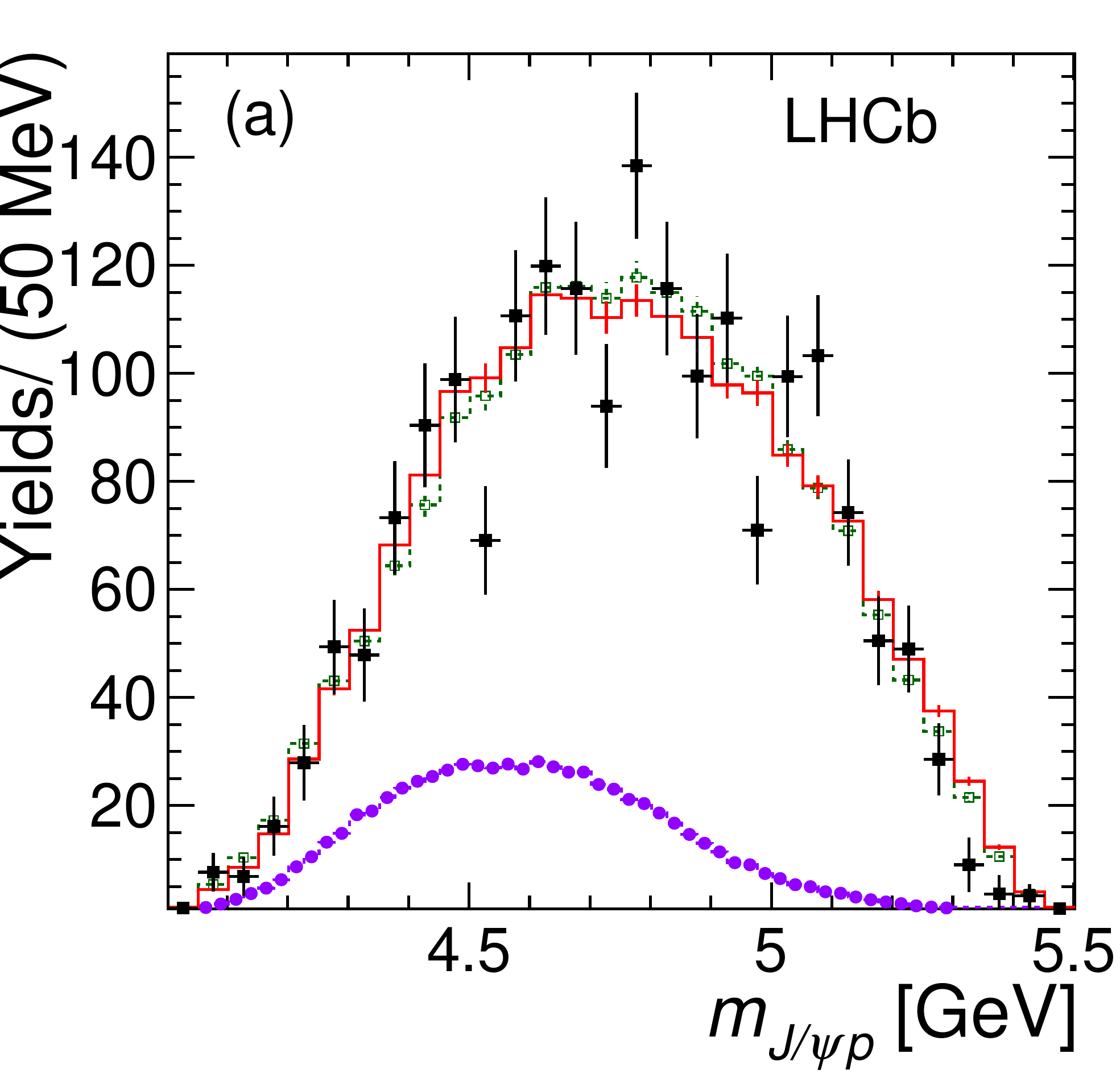}%
\includegraphics[width=0.5\textwidth]{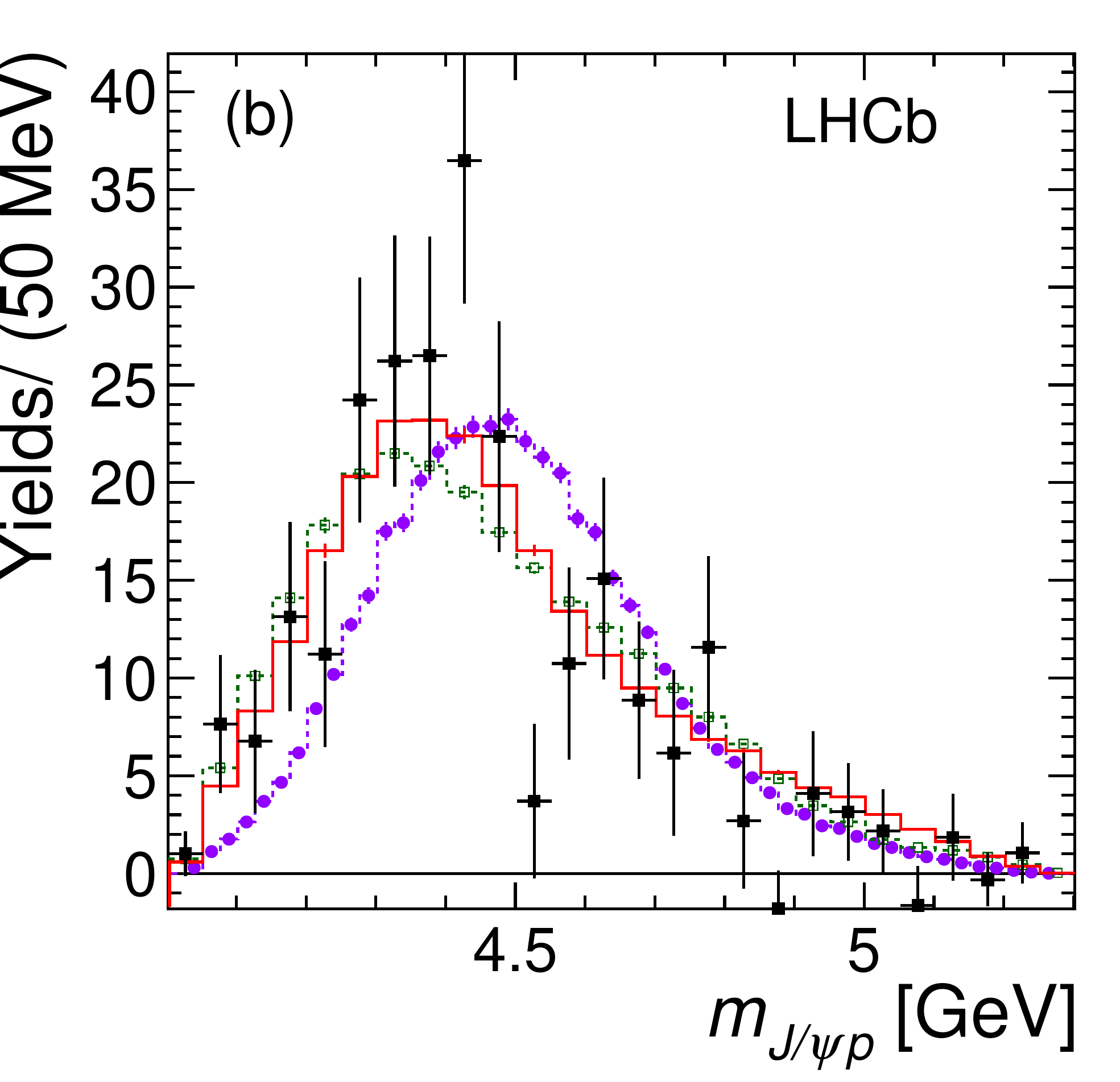}
\caption{Background-subtracted data and fit projections onto $\mjpsip$ for (a) all events and (b) the $\mppi>1.8$ \gev region. 
         See the legend and caption of Fig.~\ref{fig:mppi-zc} for a description of the components.}\label{fig:mjpsip-zc}
\end{figure*}

\begin{figure*}[!tbp]
\centering
\includegraphics[width=0.5\textwidth]{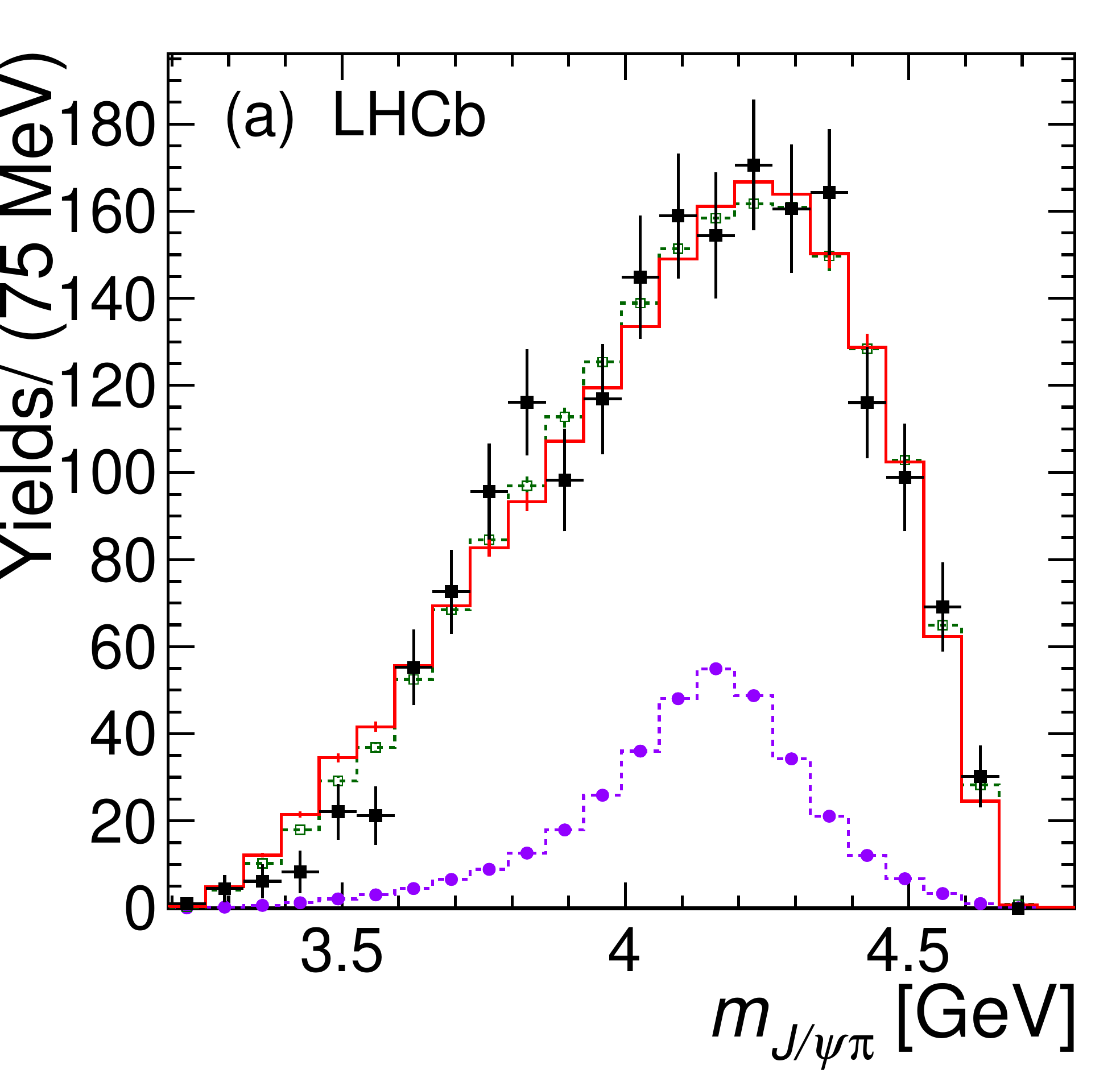}%
\includegraphics[width=0.5\textwidth]{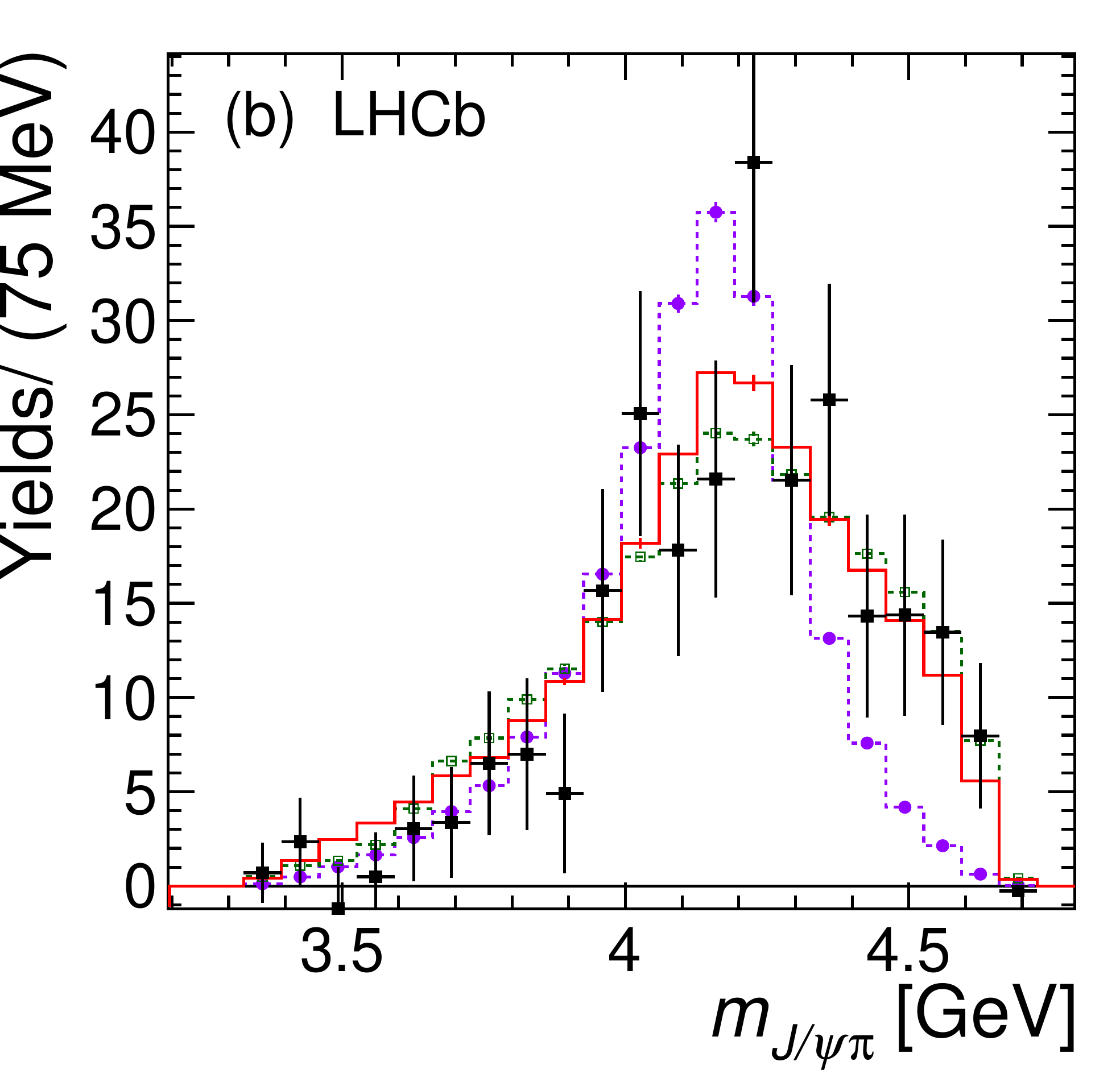}%
\caption{Background-subtracted data and fit projections onto $\mjpsipi$ for (a) all events and (b) the $\mppi>1.8$ \gev region. 
         See the legend and caption of Fig.~\ref{fig:mppi-zc} for a description of the components.}
\label{fig:mjpsipi-zc}
\end{figure*}

Figure~\ref{fig:mppi-all} shows the $\mppi$ distribution with all individual fit components overlaid.  In Fig.~\ref{fig:mppi-linear} we show the same $\mppi$ distribution but with a linear scale. 
%\subsection{Projections from the reduced model with the two $\Pcplus$ states}
The projections from the reduced model fit with the two $\Pcplus$ states are shown in Figs.~\ref{fig:mppi-2pc}--\ref{fig:mjpsipi-2pc}.
The projections from the reduced model fit with the $Z_c(4200)^-$ state are shown in Figs.~\ref{fig:mppi-zc}--\ref{fig:mjpsipi-zc}.
The models with the $P_c^+$ states describe the $m_{\jpsi p}$ 
peaking structure better than with the $Z_c(4200)^-$ alone; the $m_{\jpsi p}$ distribution is better described in Fig.~\ref{fig:mjpsip-2pc} (b) than that in Fig.~\ref{fig:mjpsip-zc} (b).  

%\subsection{Projections from the reduced model with $Z_c(4200)^-$ state}

\subsection{Fit fractions}
The fit fraction of any component $R$ is defined as $f_R = \int \left| \Mat_{R} \right|^2\,{\rm d} \Phi / \int \left| \Mat \right|^2\,{\rm d}\Phi$, 
where $\Mat_{R}$ is the matrix element, $\Mat$, with all except the $R$ amplitude terms are set to zero. The phase space volume ${\rm d}\Phi$ is equal to $p \, q \,{\rm d}\mppi \,{\rm d}\cos\theta_{\Lb} \,{\rm d}\cos\theta_{N^*}\, {\rm d}\cos\theta_{\jpsi}\, {\rm d}\phi_{\pi}\, {\rm d}\phi_{\mu}$, where $p$ is the momentum of the $p\pi$ system (\ie $N^*$) in the $\Lb$ rest frame, and $q$ is the momentum of $\pi^-$ in the $N^*$ rest frame. In Table~\ref{tab:FF}, we show the fit fractions from the ``reduced" and ``extended" model fits.
\begin{table}[t]
\centering
\caption{Fit fractions (\%) from the RM and EM model fits with statistical uncertainties only.}
\label{tab:FF}
\vspace{-0.2cm}
\begin{tabular}{lccc}
\\[-2.5ex] 
State & RM & EM  \\
\hline \\[-2.5ex] 
NR $p\pi$  & $18.6\pm3.2$  & $16.0\pm3.3$ \\
$N(1440)$ & $34.0\pm4.9$  &  $43.9\pm5.7$ \\
$N(1520)$ & $~\,7.6\pm2.2$ &  $~\,1.9\pm3.9$\\
$N(1535)$ & $25.4\pm5.9$  & $34.4\pm6.5$\\
$N(1650)$ & $10.5\pm5.1$   & $~\,9.5\pm4.1$\\
$N(1675)$ & $3.4\,_{-1.0}^{+2.2}$  & $~\,4.2\pm1.6$\\
$N(1680)$ & -  & $~\,3.0\pm1.6$\\
$N(1700)$ & -  &  $~\,1.7\pm3.0$\\
$N(1710)$ &-&  $~\,2.1\pm1.6$\\
$N(1720)$ &$3.9\,_{-1.3}^{+1.8}$ & $~\,9.6\pm3.2$\\ 
$N(1875)$ & - &  $~\,2.3\pm1.9$\\
$N(1900)$ &- &  $~\,3.0\pm1.7$ \\
$N(2190)$ &- &  $~\,0.5\pm0.4$\\  
%$N(2220)$ &- & $0.5\pm0.3$ & -\\
%$N(2250)$ &-  & $0.11\pm0.18$ & -\\
%$N(2600)$ & -  & $0.4\pm0.3$ & - \\ \hline
$N(2300)$ & -  & $~\,4.9\pm2.2$ \\
$N(2570)$ &  - & $~\,0.3\pm0.5$ \\\hline
$\ZP(4380)$& $~\,5.1\pm1.5$&$~\,4.1\pm1.7$\\
$\ZP(4450)$& $1.6\,_{-0.6}^{+0.8}$ & $1.5\,_{-0.6}^{+0.8}$\\
$\ZC(4200)$&$~\,7.7\pm2.8$ & $4.1\,_{-1.1}^{+4.3}$\\

\end{tabular}
\end{table}

\clearpage
\def\ZP{P_c}
\def\ZC{Z_c}
\def\LambdaStar{{N^*}}
\def\LambdaStarn{{N^*_{\!n}}}
\def\H{{\cal H}}
\def\F#1{\{#1\}}
\newcommand{\BA}[3]{{#1}_{{#2}}^{\,\,\F{\!#3\!}}} 
\def\Mat{\mathcal{M}}

%\section{Supplementary material for LHCb-PAPER-2016-015}
%\setcounter{section}{0}
\section{Details of the matrix element for the decay amplitude}

\subsection{Helicity formalism and notation}
\label{SUPPsec:hformalism}
 
\begin{figure}[b]
\begin{center}
\includegraphics[width=0.7\textwidth]{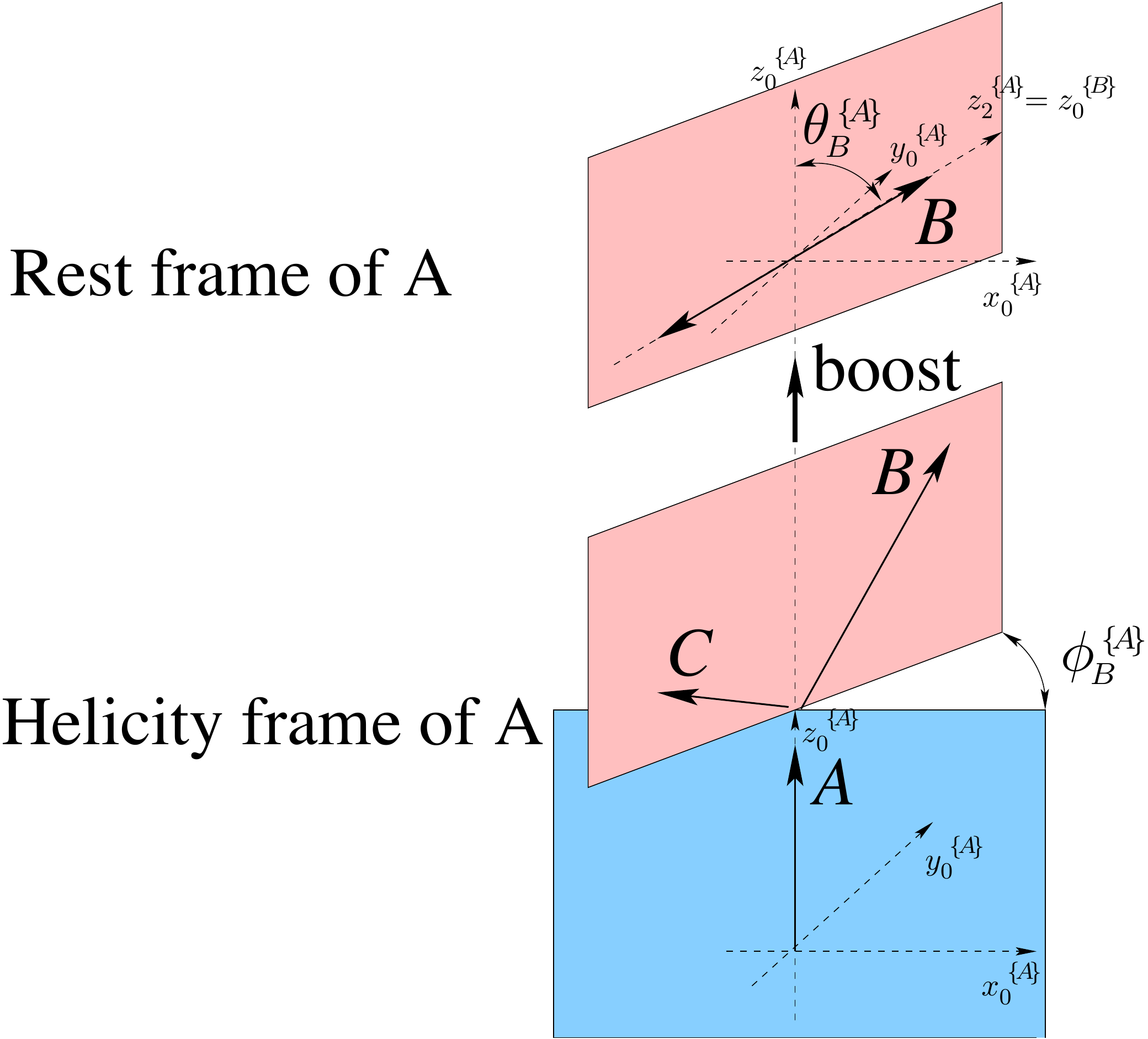}
\end{center}
\vskip -0.5cm
\caption{
Coordinate axes for the spin quantization of particle $A$ (bottom part), 
chosen to be the helicity frame of $A$ 
($\hat{z}_{0}||\vec{p}_{A}$ in the rest frame of its parent particle or in the laboratory frame),
together with the polar ($\BA{\theta}{B}{A}$) and azimuthal ($\BA{\phi}{B}{A}$) angles
of the momentum of its child $B$ in the $A$ rest frame (top part).
Notice that the directions of these coordinate axes, denoted as 
$\BA{\hat{x}}{0}{A}$, 
$\BA{\hat{y}}{0}{A}$, and 
$\BA{\hat{z}}{0}{A}$,  
do not change when boosting from the helicity frame
of $A$ to its rest frame. 
After the Euler 
rotation ${\cal R}(\alpha=\BA{\phi}{B}{A},\beta=\BA{\theta}{B}{A},\gamma=0)$
(see the text),
the rotated $z$ axis, $\BA{\hat{z}}{2}{A}$, 
is aligned with the $B$ momentum; thus the rotated coordinates
become the helicity frame of $B$. 
If $B$ has a sequential decay, then the same boost-rotation process
is repeated to define the helicity frame for its decay products.
}
\label{fig:helicitygeneric}
\end{figure}

For each two-body decay $A\to B\,C$, 
a coordinate system is set up in the rest frame of $A$, 
with $\hat{z}$ being\footnote{The ``hat'' symbol denotes a unit vector in a given direction.}
 the direction of quantization for its spin. 
We denote this coordinate system as
$(\BA{{x}}{0}{A},\BA{{y}}{0}{A},\BA{{z}}{0}{A})$,
where the superscript ``$\{A\}$'' means ``in the rest frame of $A$'', while
the subscript ``0'' means the initial coordinates.
For the first particle in the decay chain ($\Lb$), the choice of these
coordinates is arbitrary.\footnote{When designing an analysis to be sensitive (or insensitive) to a particular case of polarization, the choice is not arbitrary, but this does not change the fact that one can quantize the $\Lb$ spin along any well-defined direction.  The $\Lb$ polarization may be different for different choices.}
However, once defined, these coordinates must be used
consistently between all decay sequences described by the matrix element.
For subsequent decays, \eg\ $B\to D\,E$, the choice of these coordinates is
already fixed by the transformation from the $A$ to the $B$ rest frames, as discussed below.
Helicity is defined as the projection of the spin of the particle onto 
the direction of its momentum. When the $z$ axis coincides with the particle momentum,
we denote its spin projection onto it (\ie\ the $m_z$ quantum number) as $\lambda$.  
To use the helicity formalism, the initial coordinate system must be rotated to
align the $z$ axis with the direction of the momentum of one of the child particles, \eg\ the $B$. 
A generalized rotation operator can be formulated
in three-dimensional space, ${\cal R}(\alpha,\beta,\gamma)$, that uses Euler angles. 
Applying this operator results in a sequence of rotations: first
 by the angle $\alpha$ about the $\hat{z}_0$ axis,
followed by the angle $\beta$ about the rotated $\hat{y}_1$ axis
and then finally by the angle $\gamma$ about the rotated $\hat{z}_2$ axis.
We use a subscript denoting the axes, to specify the rotations which have been already 
performed on the coordinates.  
The spin eigenstates of particle $A$,  $|J_A,m_A\rangle$, 
in the $(\BA{x}{0}{A},\BA{y}{0}{A},\BA{z}{0}{A})$ coordinate system can be expressed in the basis of its spin eigenstates, $|J_A,m_A'\rangle$, 
in the rotated $(\BA{x}{3}{A},\BA{y}{3}{A},\BA{z}{3}{A})$ coordinate system 
with the help of Wigner's $D$-matrices
\begin{equation}
|J_A,m_A\rangle = \sum\limits_{m_A'} D^{\,J_A}_{m_A,\,m_A'}(\alpha,\beta,\gamma)^*\, |J_A,m_A'\rangle,   
\end{equation}
where
\begin{equation}
D^{\,J}_{m,\,m'}(\alpha,\beta,\gamma)^* =
\langle J,m|{\cal R}(\alpha,\beta,\gamma)|J,m'\rangle^*=
e^{i\,m\alpha}\,\,d^{\,J}_{m,m'}(\beta)\,\,e^{i\,m'\gamma},
\label{SUPeq:dmatrix}
\end{equation}
and where the small-$d$ Wigner matrix contains known functions of $\beta$ that depend on $J,m,m'$.  
To achieve the rotation of the original $\BA{\hat{z}}{0}{A}$ axis onto the $B$ momentum 
($\BA{\vec{p}}{B}{A}$), it
is sufficient to rotate by $\alpha=\BA{\phi}{B}{A}$, $\beta=\BA{\theta}{B}{A}$,
where $\BA{\phi}{B}{A}$, $\BA{\theta}{B}{A}$ are the azimuthal and polar angles of the $B$ momentum 
vector in the original coordinates \ie\ $(\BA{\hat{x}}{0}{A},\BA{\hat{y}}{0}{A},\BA{\hat{z}}{0}{A})$.
This is depicted in Fig.~\ref{fig:helicitygeneric}, for the case when the quantization axis for the
spin of $A$ is  its momentum in some other reference frame.
Since the third rotation is not necessary, we set $\gamma=0$.\footnote{An alternate convention is
to set $\gamma=-\alpha$. The two conventions lead to equivalent formulae.}
The angle $\BA{\theta}{B}{A}$ is usually called ``the $A$ helicity angle'', thus to simplify the notation
we will denote it as $\theta_A$. 
For compact notation, we will also denote $\BA{\phi}{B}{A}$ as $\phi_B$.  
These angles can be determined from\footnote{The function atan2$(x,y)$ is the $\tan^{-1}(y/x)$ function with two arguments. The purpose of using two arguments
instead of one is to gather information on the signs of the inputs in order to return the appropriate
quadrant of the computed angle.}
\begin{align}
\phi_B & =  {\rm atan2}\left( {\BA{p}{B}{A}}_{\!y},\,{\BA{p}{B}{A}}_{\!x} \right) \notag\\
       & =  {\rm atan2}\left( \BA{\hat{y}}{0}{A} \cdot\BA{\vec{p}}{B}{A} ,\, \BA{\hat{x}}{0}{A}\cdot\BA{\vec{p}}{B}{A} \right) \notag\\
       & =  {\rm atan2}\left( (\BA{\hat{z}}{0}{A} \times\BA{\hat{x}}{0}{A}) \cdot\BA{\vec{p}}{B}{A} ,\, \BA{\hat{x}}{0}{A} \cdot \BA{\vec{p}}{B}{A} \right),
   \label{SUPeq:ph}\\
\cos\theta_A & = \BA{\hat{z}}{0}{A} \cdot \BA{\hat{p}}{B}{A}. \label{SUPeq:theta} 
\end{align}

Angular momentum conservation requires $m_A'=m_B'+m_C'=\lambda_B - \lambda_C$
(since $\BA{\vec{p}}{C}{A}$ points in the opposite direction to $\BA{\hat{z}}{3}{A}$, $m_C'=-\lambda_C$).
Each two-body decay contributes a multiplicative term to the matrix element
\begin{equation}
 \H_{\lambda_B,\,\lambda_C}^{A\to B\,C}\,D^{\,J_A}_{m_A,\,\lambda_B-\lambda_C}(\phi_B,\theta_A,0)^*.
\label{SUPeq:ABCterm}
\end{equation}
The helicity couplings $\H_{\lambda_B,\,\lambda_C}^{A\to B\,C}$ are complex constants.
Their products from subsequent decays are to be determined by the fit to the data
(they represent the decay dynamics).
If the decay is strong or electromagnetic, it conserves parity which reduces the number of independent helicity couplings via the relation\begin{equation}
\H_{-\lambda_B,-\lambda_C}^{A\to B\,C} = P_A\,P_B\,P_C\,(-1)^{J_B+J_C-J_A}\, \H_{\lambda_B,\,\lambda_C}^{A\to B\,C},
\label{SUPeq:parity}
\end{equation}
where $P$ stands for the intrinsic parity of a particle. 

After multiplying terms given by Eq.~(\ref{SUPeq:ABCterm}) for all decays in the decay sequence, they must be summed up coherently over the helicity states of intermediate particles, and incoherently over  the helicity states of the initial and final-state particles. Possible helicity values of $B$ and $C$ particles are constrained by 
$|\lambda_B|\le J_B$, $|\lambda_C|\le J_C$ and $|\lambda_B-\lambda_C|\le J_A$.

When dealing with the subsequent decay of the child, $B\to D\, E$, 
four-vectors of all particles must be first Lorentz boosted to the rest frame of $B$, along 
the $\BA{\vec{p}}{B}{A}$ \ie\ $\BA{\hat{z}}{3}{A}$ direction (this is the $z$ axis in the rest frame of 
$A$ after the Euler rotations; we use the subscript ``3'' for the number of rotations performed on the
coordinates, because of the three Euler angles, however, since we use the $\gamma=0$ convention 
these coordinates are the same as after the first two rotations). 
This is visualized in Fig.~\ref{fig:helicitygeneric}, 
with $B\to D\, E$ particle labels replaced by $A\to B\, C$ labels. 
This transformation does not change vectors that are perpendicular to the boost direction.
The transformed coordinates become the initial coordinate system quantizing the spin of $B$ in its
rest frame,
\begin{align}
\BA{\hat{x}}{0}{B} & =\BA{\hat{x}}{3}{A}, \notag\\  
\BA{\hat{y}}{0}{B} & =\BA{\hat{y}}{3}{A}, \notag\\  
\BA{\hat{z}}{0}{B} & =\BA{\hat{z}}{3}{A}.       
\label{SUPeq:xbfromxa}
\end{align}
The processes of rotation and subsequent boosting can be repeated until the final-state particles are reached.
In practice, there are two equivalent ways to determine the $\BA{\hat{z}}{0}{B}$ direction.
Using Eq.~(\ref{SUPeq:xbfromxa}) we can set it to the direction of  the $B$ momentum in the $A$ rest frame
\begin{equation}
\BA{\hat{z}}{0}{B} =\BA{\hat{z}}{3}{A} = \BA{\hat{p}}{B}{A}.
\label{SUPeq:zboost}
\end{equation}
Alternatively, we can make use of the fact that $B$ and $C$ are 
back-to-back in the rest frame of $A$, $\BA{\vec{p}}{C}{A}= - \BA{\vec{p}}{B}{A}$.
Since the momentum of $C$ is antiparallel to the boost direction from the $A$ to $B$ rest frames,
the $C$ momentum in the $B$ rest frame will be different, but it will still be antiparallel to this
boost direction
\begin{equation}
\BA{\hat{z}}{0}{B} = - \BA{\hat{p}}{C}{B}.
\label{SUPeq:zrecoil}
\end{equation}
To determine $\BA{\hat{x}}{0}{B}$
from Eq.~(\ref{SUPeq:xbfromxa}), we need to find  $\BA{\hat{x}}{3}{A}$.
After the first rotation by $\phi_B$ about $\BA{\hat{z}}{0}{A}$, 
the $\BA{\hat{x}}{1}{A}$ axis is along the component of 
$\BA{\vec{p}}{B}{A}$ which is perpendicular to the $\BA{\hat{z}}{0}{A}$ axis
\begin{align}
\BA{\vec{a}}{B\perp z_0}{A} & \equiv (\BA{\vec{p}}{B}{A})_{\perp \BA{\hat{z}}{0}{A}}                   
                    = {\BA{\vec{p}}{B}{A}} - (\BA{\vec{p}}{B}{A})_{|| \BA{\hat{z}}{0}{A}}, \notag\\
                   & = {\BA{\vec{p}}{B}{A}} - ({\BA{\vec{p}}{B}{A}}\cdot\BA{\hat{z}}{0}{A})\,\BA{\hat{z}}{0}{A}, \notag\\
%\BA{\hat{x}}{1}{A} & = {\BA{\hat{p}}{B}{A}}~_{\perp \BA{\hat{z}}{0}{A}} 
%                     = \frac{{\BA{\vec{p}}{B}{A}}~_{\perp \BA{\hat{z}}{0}{A}}}{
%                             |\,{\BA{\vec{p}}{B}{A}}~_{\perp \BA{\hat{z}}{0}{A}}\,|}
\BA{\hat{x}}{1}{A} & = \BA{\hat{a}}{B\perp z_0}{A} = \frac{\BA{\vec{a}}{B\perp z_0}{A}}{|\,\BA{\vec{a}}{B\perp z_0}{A}\,|}.
\label{SUPeq:xonea}
\end{align}
After the second rotation by $\theta_A$ about $\BA{\hat{y}}{1}{A}$,
$\BA{\hat{z}}{2}{A}\equiv \BA{\hat{z}}{3}{A} = \BA{\hat{p}}{B}{A}$, and 
$\BA{\hat{x}}{2}{A}= \BA{\hat{x}}{3}{A}$ is antiparallel to 
the component of the $\BA{\hat{z}}{0}{A}$ vector that is perpendicular to 
the new $z$ axis \ie\ $\BA{\hat{p}}{B}{A}$. Thus
\begin{align}
\BA{\vec{a}}{z_0\perp B}{A} & \equiv
(\BA{\hat{z}}{0}{A})_{\perp \BA{\vec{p}}{B}{A}}                   
                    = {\BA{\hat{z}}{0}{A}} - ({\BA{\hat{z}}{0}{A}}\cdot\BA{\hat{p}}{B}{A})\,\BA{\hat{p}}{B}{A}, \notag\\
\BA{\hat{x}}{0}{B} &=
\BA{\hat{x}}{3}{A}  = \,-\,\, \BA{\hat{a}}{z_0\perp B}{A} = \,-\,\,\frac{\BA{\vec{a}}{z_0\perp B}{A}}{|\,\BA{\vec{a}}{z_0\perp B}{A}\,|}.
%\frac{{\BA{\hat{z}}{0}{A}}~_{\perp \BA{\vec{p}}{B}{A}}}{
%                               |\,{\BA{\hat{z}}{0}{A}}~_{\perp \BA{\vec{p}}{B}{A}}\,|}.                      
\label{SUPeq:xaxis}
\end{align}
Then we obtain $\BA{\hat{y}}{0}{B}=\BA{\hat{z}}{0}{B}\times\BA{\hat{x}}{0}{B}$. 

If $C$ also decays, $C\to F\, G$, then the coordinates for the quantization of $C$ spin
in the $C$ rest frame are defined by
\begin{align}
\BA{\hat{z}}{0}{C} & = - \BA{\hat{z}}{3}{A} = \BA{\hat{p}}{C}{A} = - \BA{\hat{p}}{B}{C}, \\        
\BA{\hat{x}}{0}{C} & =  \BA{\hat{x}}{3}{A} = \,-\, \BA{\hat{a}}{z_0\perp B}{A} = + \BA{\hat{a}}{z_0\perp C}{A}, \label{SUPeq:xcaxis} \\  
\BA{\hat{y}}{0}{C} & = \BA{\hat{z}}{0}{C}\times\BA{\hat{x}}{0}{C}, 
\label{SUPeq:xcfromxa}
\end{align}
\ie\ the $z$ axis is reflected compared to the system used for the decay of particle $B$
(it must point in the direction of $C$ momentum in the $A$ rest frame),
but the $x$ axis is kept the same, since we chose particle $B$ for the rotation used 
in Eq.~(\ref{SUPeq:ABCterm}).

\subsection{Matrix element for the $\ZC^-$ decay chain}
\label{Supp:ZCME}

\begin{figure}[b]
\begin{center}
\hbox{\includegraphics[width=.95\textwidth]{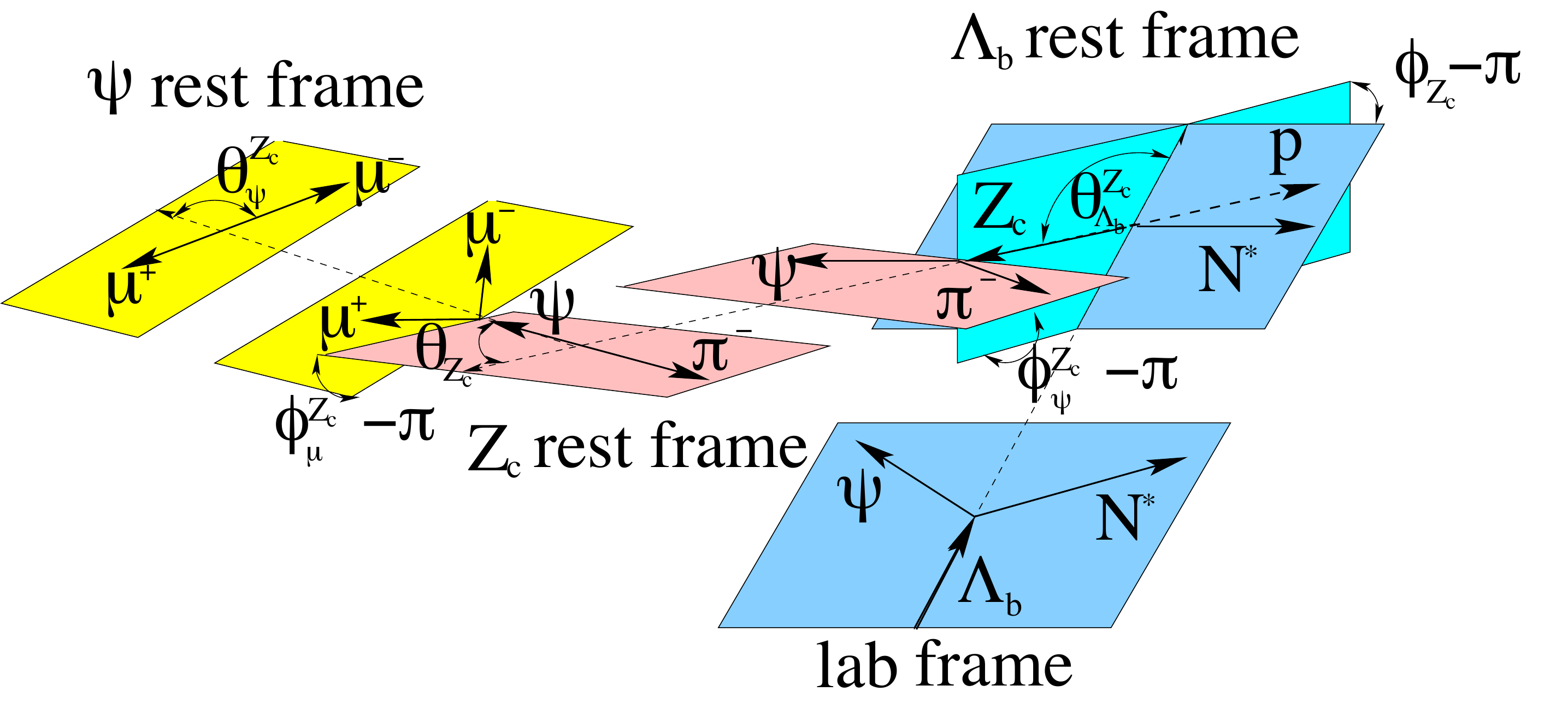}\hskip-4cm\quad}
\end{center}
\vskip -0.8cm
\caption{
Definition of the decay angles in the $\ZC^-$ decay chain. 
}
\label{fig:helicityzc}
\end{figure} 
%_f
The decay matrix elements for the two interfering decay chains, 
$\Lb \to \jpsi N^*$, $N^* \to p \pim$ 
and $\Lb\to \Pcplus \pim$, $\Pcplus \to \jpsi p$ 
with $\jpsi \to \mup \mu^-$ in both cases, 
are identical to those used in the $\Lb \to \jpsi p K^-$ analysis \cite{LHCb-PAPER-2015-029}, 
with $K^-$ and ${\Lz^*}$ replaced by $\pi^-$ and $N^*$. %Next we discuss  the additional interfering decay chain, $\Lb \to \Zcminus p$, $\Zcminus \to \jpsi \pim$.
We now turn to the discussion of  the additional interfering decay chain,  $\Lb\to {\ZC}_f p$, ${\ZC}_f\to\psi \pi^-$ decays (denoting $\jpsi$ as $\psi$), in which we 
allow more than one tetraquark state, $f=1, 2, \ldots\,\,$. 
Superscripts containing the $\ZC$ decay chain name without curly brackets, \eg\ $\phi^{\,\ZC}$, 
 will denote quantities belonging to this decay chain and should not
be confused with the superscript ``$\{\ZC\}$'' denoting the $\ZC$ rest frame, \eg\ $\phi^{\,\,\F{\ZC}}$.
With only a few exceptions, we omit the $\LambdaStar$ decay chain label. The angular calculations for the $\ZC^-$ decay chain are analogous to that for $P_c^+$ by interchange of $p$ and $\pi^-$, except for the angles to align the proton helicity.

The weak decay $\Lb\to{\ZC}_f p$ 
is described by the term,
\begin{equation}
\label{SUPeq:lbtozk}
\H^{\Lb\to {\ZC}_f p}_{\lambda_{\ZC},\lambda_{p}^{\ZC}} \,\,
D^{\,\,\frac{1}{2}}_{\lambda_{\Lb},\,\lambda_{\ZC}-\lambda_{p}^{\ZC}}(
\phi_{\ZC},\theta_{\Lb}^{\ZC},0)^*,
\end{equation}
where $\H^{\Lb\to {\ZC}_f p}_{\lambda_{\ZC},\lambda_{p}^{\ZC}}$ 
are resonance (\ie\ $f$) dependent helicity couplings. 
As for $|\lambda_{\ZC}-\lambda_{p}^{\ZC}|\le\frac{1}{2}$, 
there are two different helicity couplings for $J_{\ZC}=0$, and four for higher spin.
%The helicity of the tetraquark state, $\lambda_{\ZC}$, can take
%values of $\pm\frac{1}{2}$ independently of its spin, 
%$J_{{\ZC}_f}=\frac{1}{2}, \frac{3}{2}, \ldots\,\,$.
 The above mentioned $\phi_{\ZC}$, $\theta_{\Lb}^{\ZC}$ symbols refer to 
the azimuthal and polar angles of $\ZC$ in the $\Lb$ rest frame
(see Fig.~\ref{fig:helicityzc}).

With the direction of $\Lb$ in the lab frame $\BA{\hat{p}}{\Lb}{{\rm lab}}$, and the direction of $\ZC$ in the $\Lb$ rest frame,
the $\Lb$ helicity angle in the $\ZC$ decay chain can be calculated as,
\begin{equation}
\label{SUPeq:lbhelzc}
\cos\theta_{\Lb}^{\ZC}= \BA{\hat{p}}{\Lb}{{\rm lab}} \cdot \BA{\hat{p}}{\ZC}{\Lb}.
\end{equation}

The $\phi_{\ZC}$ angle cannot be set to zero, 
since we have already defined the $\BA{\hat{x}}{0}{\Lb}$ axis in the $\Lb$ rest frame 
by the $\phi_{\LambdaStar}=0$ convention: 
\begin{align}
\BA{\vec{a}}{\LambdaStar\perp z_0}{\Lb}
                   & = \BA{\vec{p}}{\LambdaStar}{\Lb} - (\BA{\vec{p}}{\LambdaStar}{\Lb}\cdot\BA{\hat{p}}{\Lb}{{\rm lab}})\,\BA{\hat{p}}{\Lb}{{\rm lab}}, \notag\\
\BA{\hat{x}}{0}{\Lb} 
                   & =  \frac{\BA{\vec{a}}{\LambdaStar\perp z_0}{\Lb}}{
                         |\,\BA{\vec{a}}{\LambdaStar\perp z_0}{\Lb}\,|}.
\label{SUPeq:xlambdabzc}
\end{align}
The $\phi_{\ZC}$ angle can be determined in the $\Lb$ rest frame from
\begin{equation}
\label{SUPeq:phzc}
\phi_{\ZC} = {\rm atan2} \left( (\BA{\hat{p}}{\Lb}{{\rm lab}}\times\BA{\hat{x}}{0}{\Lb})\cdot\BA{\hat{p}}{\ZC}{\Lb},\, \BA{\hat{x}}{0}{\Lb}\cdot\BA{\hat{p}}{\ZC}{\Lb} \right).
\end{equation}

The strong decay ${\ZC}_f\to \psi \pi^-$ is described by a term
\begin{equation}
\label{SUPeq:zctopsip}
\H^{{\ZC}_f\to \psi \pi}_{\lambda_\psi^{\ZC}} \,\,
D^{\,\,J_{{\ZC}_f}}_{\lambda_{\ZC},\,\lambda_\psi^{\ZC}}(
\phi_{\psi}^{\ZC},\theta_{\ZC},0)^*\,\,
R_{{\ZC}_f}(M_{\psi \pi}),
\end{equation}
where $\phi_{\psi}^{\ZC},\theta_{\ZC}$ are the azimuthal and polar angles of the $\psi$ in the $\ZC$ rest frame 
(see Fig.~\ref{fig:helicityzc}).
%They are defined analogously to Eqs.~(\ref{SUPeq:thl})$-$(\ref{SUPeq:phk}).  
The $\BA{\hat{z}}{0}{\ZC}$ direction is defined by the boost direction from the $\Lb$ rest frame, which coincides
with the $-\BA{\vec{p}}{p}{\ZC}$ direction. This leads to
\begin{equation}
\cos\theta_{\ZC}= - \BA{\hat{p}}{p}{\ZC} \cdot \BA{\hat{p}}{\psi}{\ZC}.
\end{equation}
%\end{align}
The azimuthal angle of the $\psi$ can now be determined in the $\ZC$ rest frame (see Fig.~\ref{fig:helicityzc}) from
\begin{equation}
\label{SUPeq:phpsi}
\phi_\psi^{\ZC} = {\rm atan2} \left( -(\BA{\hat{p}}{p}{\ZC}\times\BA{\hat{x}}{0}{\ZC})\cdot\BA{\hat{p}}{\psi}{\ZC},\, \BA{\hat{x}}{0}{\ZC}\cdot\BA{\hat{p}}{\psi}{\ZC} \right).
\end{equation}

The $\BA{\hat{x}}{0}{\ZC}$ direction is defined by the convention that we used in the
$\Lb$ rest frame. Thus, we have
\begin{align}
\BA{\vec{a}}{z_0\perp\ZC}{\Lb}
                   & = \BA{\hat{p}}{\Lb}{{\rm lab}} - (\BA{\hat{p}}{\Lb}{{\rm lab}}\cdot\BA{\hat{p}}{\ZC}{\Lb})\,\BA{\hat{p}}{\ZC}{\Lb} ,\notag\\
\BA{\hat{x}}{0}{\ZC} 
                   & =\, -\,\, \frac{\BA{\vec{a}}{z_0\perp\ZC}{\Lb}}{
                               |\,\BA{\vec{a}}{z_0\perp\ZC}{\Lb}\,|}.
\label{SUPeq:xzc}
\end{align}  %%SS input
Again, the $\psi$ and $p$ helicities are labeled as $\lambda_\psi^{\ZC}$ and $\lambda_p^{\ZC}$,
with the $\ZC$ superscript to make it clear that the spin quantization axes are different than in the $\LambdaStar$ decay chain.
Since the $\psi$ is an intermediate particle, this has no consequences after we sum (coherently) over $\lambda_\psi^{\ZC}=-1,0,+1$.
The proton, however, is a final-state particle. Before the $\ZC$ terms in the matrix element can be added coherently to the $\LambdaStar$ terms,
the $\lambda_p^{\ZC}$ states must be rotated to $\lambda_p$ states (defined in the $\LambdaStar$ decay chain).
The proton helicity axes are different, since the proton comes from a decay of different particles in the two decay sequences, 
the $\LambdaStar$ and $\Lb$. 
The quantization axes are along the proton direction in the $\LambdaStar$ and the $\Lb$ rest frames, thus antiparallel 
to the particles recoiling against the proton: the $\pi^-$ and $\ZC$, respectively. These directions are preserved when 
boosting to the proton rest frame.
Thus, the polar angle between the two proton quantization  axes ($\theta_p^{\ZC}$) can be determined from
the opening angle between the $\pi^-$ and $\ZC$ mesons in the $p$ rest frame, 
\begin{equation}
\cos\theta_p^{\ZC}=\BA{\hat{p}}{\pi}{p}\cdot\BA{\hat{p}}{\ZC}{p}.
\end{equation}
The dot product above must be calculated by operating on the $\BA{\vec{p}}{\pi}{p}$ and $\BA{\vec{p}}{\ZC}{p}$ vectors 
in the proton rest frame obtained by  the same sequence of  boost transformations, either according to the $\LambdaStar$ or 
$\ZC$ decay chains, or even by a direct boost transformation 
from the {\rm lab} frame.
%This may appear confusing since the formula itself corrects for the different boost sequences in the two decay chains. 
%However, the different boost sequences are already encoded in the choice of particle momenta that enter the formula.    
%
%\begin{figure}[t!]
%\begin{center}
%%\includegraphics[width=\textwidth]{MatrixElement-theta_p-Zc}
%\end{center}
%\vskip-5mm
%\caption{
%Definition of the $\theta_p$ angle.
%}
%\label{fig:helicityp}
%\end{figure} 

Unlike in the $P_c$ decay chain, the azimuthal angle ($\alpha_p^{\ZC}$) aligning the two proton helicity frames is not zero. The angle can be determined from 

\begin{equation}
\label{SUPeq:alphap}
\alpha_p^{\ZC}={\rm atan2} \left( (\BA{\hat{z}}{0}{p}~^{\Lb}\times\BA{\hat{x}}{0}{p}~^{\Lb})\cdot\BA{\hat{x}}{0}{p}~^{\LambdaStar}\,, \BA{\hat{x}}{0}{p}~^{\Lb}\cdot\BA{\hat{x}}{0}{p}~^{\LambdaStar}\right),
\end{equation}
where all vectors are in the $p$ rest frame, $\BA{\hat{x}}{0}{p}~^{\LambdaStar}$ is the direction of the $x$ axis when boosting from the $\LambdaStar$ rest frame, $\BA{\hat{x}}{0}{p}~^{\Lb}$ and $\BA{\hat{z}}{0}{p}~^{\Lb}$ are the directions of the $x$ and $z$ axes when boosting from the $\Lambda_b^0$ rest frame. From Eq.~(\ref{SUPeq:zrecoil}), $\BA{\hat{z}}{0}{p}~^{\Lb}=-\BA{\hat{p}}{\ZC}{p}$. Direction of $\BA{\hat{x}}{0}{p}~^{\Lb}$ is given by Eq.(\ref{SUPeq:xaxis}) with ${\BA{\hat{z}}{0}{\Lb}} ={\BA{\hat{p}}{\Lb}{\rm lab}} $

\begin{align}
\BA{\vec{a}}{z_0\perp \ZC}{\Lb} &                 
                    = {\BA{\hat{p}}{\Lb}{\rm lab}}  - ({\BA{\hat{p}}{\Lb}{\rm lab}} \cdot\BA{\hat{p}}{\ZC}{\Lb})\,\BA{\hat{p}}{\ZC}{\Lb}, \notag\\
\BA{\hat{x}}{0}{p}~^{\Lb} & =  \frac{\BA{\vec{a}}{z_0\perp \ZC}{\Lb}}{|\,\BA{\vec{a}}{z_0\perp \ZC}{\Lb}\,|}.
\end{align}

Therefore, the relation between $\lambda_p$ and $\lambda_p^{\ZC}$ states is
\begin{equation}
|\lambda_p\rangle = \sum\limits_{\lambda_p^{\ZC}} D^{\,\,J_p}_{\lambda_p^{\ZC},\,\lambda_p}(\alpha_p^{\ZC},\theta_p^{\ZC},0)^* |\lambda_p^{\ZC}\rangle
            = \sum\limits_{\lambda_p^{\ZC}} e^{i\lambda_p^{\ZC}\alpha_p^{\ZC}}d^{\,\,J_p}_{\lambda_p^{\ZC},\,\lambda_p}(\theta_p^{\ZC}) |\lambda_p^{\ZC}\rangle.
\end{equation}
Thus, the term given by Eq.~(\ref{SUPeq:zctopsip}) must be preceded by
\begin{equation}
\label{SUPeq:prot2}
\sum\limits_{\lambda_p^{\ZC}=\pm\frac{1}{2}} e^{i\lambda_p^{\ZC}\alpha_p^{\ZC}}d^{\,\,J_p}_{\lambda_p^{\ZC},\,\lambda_p}(\theta_p^{\ZC}).
\end{equation}

Parity conservation in ${\ZC}_f\to \psi \pi^-$ decays leads to the following relation
\begin{align}
\H^{{\ZC}_f\to \psi \pi}_{-\lambda_\psi^{\ZC}} 
& = P_\psi\,P_{\pi}\,P_{{\ZC}_f}\,(-1)^{J_\psi+J_K-J_{{\ZC}_f}}\,
   \H^{{\ZC}_f\to \psi \pi}_{\lambda_\psi^{\ZC}} \notag\\
& = P_{{\ZC}_f}\, (-1)^{1-J_{{\ZC}_f}}\,
   \H^{{\ZC}_f\to \psi \pi}_{\lambda_\psi^{\ZC}}, 
\label{SUPeq:zcparity}
\end{align}
where $P_{{\ZC}_f}$ is the parity of the ${\ZC}_f$ state.
Then the number of independent helicity couplings to be determined from the data
is reduced to two for $J_{{\ZC}_f}\ge1$ and remains equal to unity for $J_{{\ZC}_f}=0$.
Since the helicity couplings enter the matrix element formula as a product,  
$\H^{\Lb\to {\ZC}_f p}_{\lambda_{\ZC},\,\lambda_p^{\ZC}}\,\H^{{\ZC}_f\to \psi \pi}_{\lambda_\psi^{\ZC}}$, the
relative magnitude and phase of these two sets must be fixed by a convention.
For example, $\H^{{\ZC}_f\to \psi \pi}_{\lambda_\psi^{\ZC}=0}$ can be set to $(1,0)$ for
every ${\ZC}_f$ resonance, in which case
$\H^{{\ZC}_f\to \psi \pi}_{\lambda_\psi^{\ZC}=1}$ develops a meaning of the complex ratio of
$\H^{{\ZC}_f\to \psi \pi}_{\lambda_\psi^{\ZC}=1}/\H^{{\ZC}_f\to \psi \pi}_{\lambda_\psi^{\ZC}=0}$,
while all $\H^{\Lb\to {\ZC}_f p}_{\lambda_{\ZC},\,\lambda_p^{\ZC}}$ couplings should have both
real and imaginary parts free in the fit. 

The term $R_{{\ZC}_f}(M_{\psi \pi})$ in Eq. (\ref{SUPeq:zctopsip}) describes the $\psi \pi$  invariant mass distribution of the 
${\ZC}_f$ resonance. 
%and is given by Eq.~(\ref{SUPeq:resshape}) after appropriate substitutions.
Angular momentum conservation restricts ${\rm max}(J_{{\ZC}_f}-1,0)\le L_{\Lb}^{{\ZC}_f}\le J_{{\ZC}_f}+1$ in $\Lb \to {\ZC}_f p$ decays.
Angular momentum conservation also imposes 
${\rm max}(|J_{{\ZC}_f}-1|\,, \,0)\le L_{{\ZC}_f} \le J_{{\ZC}_f}+1$,
which is further restricted by the parity conservation in the ${\ZC}_f$ decays,
$P_{{\ZC}_f}=(-1)^{L_{{\ZC}_f}}$.
We assume the minimal values of $L_{\Lb}^{{\ZC}_f}$ and of $L_{{\ZC}_f}$ in $R_{{\ZC}_f}(m_{\psi \pi})$.

The electromagnetic decay $\psi\to\mu^+\mu^-$ in the $\ZC$ decay chain contributes a term
\begin{equation}
\label{SUPeq:psitommz}
D^{\,\,1}_{\lambda_{\psi}^{\ZC},\,\Delta\lambda_\mu^{\ZC}}(
\phi_{\mu}^{\ZC},\theta_{\psi}^{\ZC},0)^*.
\end{equation}
%which is the same as Eq.~(\ref{SUPeq:psitomm}), except that since the $\psi$ meson comes from the decay of different particles 
%in the two decay chains, 
The azimuthal and polar angle of the muon in the $\psi$ rest frame, 
$\phi_{\mu}^{\ZC}$, $\theta_{\psi}^{\ZC}$, are different from 
$\phi_{\mu}$, $\theta_{\psi}$ introduced in the $\LambdaStar$ decay chain.
The $\psi$ helicity axis is along the boost direction from the $\ZC$ to the $\psi$ rest frames, which 
is given by
\begin{equation}
\BA{\hat{z}}{0}{\psi}~^{\ZC}  =  \, - \, \BA{\hat{p}}{\pi}{\psi}, 
\end{equation}
and so
\begin{equation}
\cos\theta_{\psi}^{\ZC}  =  - \BA{\hat{p}}{\pi}{\psi} \cdot \BA{\hat{p}}{\mu}{\psi}. \label{SUPeq:psihelthz}
\end{equation}
The $x$ axis is inherited from the $\ZC$ rest frame (Eq.~(\ref{SUPeq:xaxis})),
\begin{align}
\BA{\vec{a}}{z_0\perp \psi}{\ZC} & 
                    = - {\BA{\vec{p}}{p}{\ZC}} + ({\BA{\vec{p}}{p}{\ZC}}\cdot\BA{\hat{p}}{\psi}{\ZC})\,\BA{\hat{p}}{\psi}{\ZC} \notag\\
\BA{\hat{x}}{0}{\psi}~^{\ZC} =
\BA{\hat{x}}{3}{\ZC} & = \,-\,\,\frac{\BA{\vec{a}}{z_0\perp \psi}{\ZC}}{|\,\BA{\vec{a}}{z_0\perp \psi}{\ZC}\,|},
\label{SUPeq:xpsiz}
\end{align}
which leads to
\begin{equation}
\label{SUPeq:psihelphz} 
\phi_\mu^{\ZC}  =  {\rm atan2} \left( -(\BA{\hat{p}}{\pi}{\psi}\times\BA{\hat{x}}{0}{\psi}~^{\ZC})\cdot\BA{\hat{p}}{\mu}{\psi},\, 
                                        \BA{\hat{x}}{0}{\psi}~^{\ZC}\cdot\BA{\hat{p}}{\mu}{\psi} \right).
\end{equation}
  
The azimuthal angle $\alpha_{\mu}^{\ZC}$ is defined by 
\begin{equation}
\label{SUPeq:alphamu2}
\alpha_{\mu}^{\ZC} = {\rm atan2} \left( (\BA{\hat{z}}{3}{\psi}~^{\ZC}\times\BA{\hat{x}}{3}{\psi}~^{\ZC})\cdot\BA{\hat{x}}{3}{\psi}~^{\LambdaStar},\, 
                                   \BA{\hat{x}}{3}{\psi}~^{\ZC}\cdot\BA{\hat{x}}{3}{\psi}~^{\LambdaStar} \right),
\end{equation}
where
$\BA{\hat{z}}{3}{\psi}~^{\ZC}=\BA{\hat{p}}{\mu}{\psi}~^{\ZC}$,
and from Eq.~(\ref{SUPeq:xaxis})
\begin{align}
\BA{\hat{x}}{3}{\psi}~^{\ZC} & = \, - \, \BA{\hat{a}}{z_0\perp\mu}{\psi}~^{\ZC}, \\
\BA{\vec{a}}{z_0\perp\mu}{\psi}~^{\ZC} & 
   = \, - \, \BA{\hat{p}}{\pi}{\psi} + (\BA{\hat{p}}{\pi}{\psi}\cdot\BA{\hat{p}}{\mu}{\psi})\,\BA{\hat{p}}{\mu}{\psi},
\end{align}
as well as
\begin{align}
\BA{\hat{x}}{3}{\psi}~^{\LambdaStar} & = \, - \, \BA{\hat{a}}{z_0\perp\mu}{\psi}~^{\LambdaStar}, \\
\BA{\vec{a}}{z_0\perp\mu}{\psi}~^{\LambdaStar} & 
   = \, - \, \BA{\hat{p}}{\LambdaStar}{\psi} + (\BA{\hat{p}}{\LambdaStar}{\psi}\cdot\BA{\hat{p}}{\mu}{\psi})\,\BA{\hat{p}}{\mu}{\psi}.
\end{align}

Collecting terms from the three subsequent decays in the $\ZC$ chain together, 
\begin{align}  
\Mat_{\lambda_{\Lb},\,\lambda_p^{\ZC},\,\Delta\lambda_\mu^{\ZC}}^{\,\,\ZC} =  
&
  e^{i\,\lambda_{\Lb}\phi_{\ZC}}
  \sum\limits_{f} 
     R_{{\ZC}_f}(M_{\psi \pi}) \,
     \sum\limits_{\lambda_\psi^{\ZC}}
        e^{i\,\lambda_\psi^{\ZC}\phi_{\mu}^{\ZC}} \,\,
        d^{\,\,1}_{\lambda_\psi^{\ZC},\,\Delta\lambda_\mu}(\theta_\psi^{\ZC}) \,\,
	\notag \\
& ~~\times \sum\limits_{\lambda_{\ZC}} 
        \H^{\Lb\to {\ZC}_f p}_{\lambda_{\ZC},\,\lambda_{p}^{\ZC}} \,\,
        e^{i\,\lambda_{\ZC} \phi_\psi^{\ZC}} \,\,
        d^{\,\,\frac{1}{2}}_{\lambda_{\Lb},\,\lambda_{\ZC}-\lambda_p^{\ZC}}(\theta_{\Lb}^{\ZC}) 
        \H^{{\ZC}_f\to\psi \pi}_{\lambda_{\psi}^{\ZC}}\,\,
        d^{\,\,J_{{\ZC}_f}}_{\lambda_{\ZC},\,\lambda_{\psi}^{\ZC}}(\theta_{\ZC}),
\label{SUPeq:Pc_matrixelement_partial}
\end{align}
and adding them coherently to the $\LambdaStar$ and the $\ZP$ matrix elements, 
via appropriate relations of $|\lambda_p\rangle|\lambda_{\mu^+}\rangle|\lambda_{\mu^-}\rangle$ to 
$|\lambda_p^{\ZP}\rangle|\lambda_{\mu^+}^{\ZP}\rangle|\lambda_{\mu^-}^{\ZP}\rangle$ and 
$|\lambda_p^{\ZC}\rangle|\lambda_{\mu^+}^{\ZC}\rangle|\lambda_{\mu^-}^{\ZC}\rangle$ states as discussed above,
leads to the final matrix element squared
\begin{align} 
\left| \Mat \right|^2 = 
\sum\limits_{\lambda_{\Lb}=\pm\frac{1}{2}}
\sum\limits_{\lambda_{p}=\pm\frac{1}{2}}
\sum\limits_{\Delta\lambda_{\mu}=\pm1}
&\left|
\Mat_{\lambda_{\Lb},\,\lambda_p,\,\Delta\lambda_\mu}^{\LambdaStar} 
+ 
e^{i\,{\Delta\lambda_\mu}\alpha_{\mu}}\,
\sum\limits_{\lambda_p^{\ZP}} 
d^{\,\,\frac{1}{2}}_{\lambda_p^{\ZP},\,\lambda_p}(\theta_p)\,
\Mat_{\lambda_{\Lb},\,\lambda_p^{\ZP},\,\Delta\lambda_\mu}^{\ZP}  \right. \notag \\ 
&+ \left.
e^{i\,{\Delta\lambda_\mu}\alpha^{\ZC}_{\mu}}\,
\sum\limits_{\lambda_p^{\ZC}} e^{i\lambda_p^{\ZC}\alpha_p^{\ZC}}
d^{\,\,\frac{1}{2}}_{\lambda_p^{\ZC},\,\lambda_p}(\theta^{\ZC}_p)\,
\Mat_{\lambda_{\Lb},\,\lambda_p^{\ZC},\,\Delta\lambda_\mu}^{\ZC} 
\right|^2.
\label{SUPPeq:total_matrixelement}
\end{align}  
%where we set $\Pol=0$. 
%As a cross-check, $\jpsi p K^-$ study showed the fitted value yileds a value consistent 
%with zero, $\Pol=(-2.0\pm2.3)\%$ (statistical error only)~\cite{LHCb-PAPER-2015-029}. 

Assuming approximate $\CP$ symmetry, the helicity couplings for $\Lb$ and $\Lbbar$ can be made equal, but the calculation 
of the angles requires some care, since  parity ($P$) conservation does not change polar (\ie helicity) angles, but does change azimuthal angles. 
Thus, not only must $\vec{p}_{\mu^+}$ be used instead of $\vec{p}_{\mu^-}$ for $\Lbbar$ candidates (with $\pi^+$ and $\bar p$ 
in the final state) in Eqs.~(\ref{SUPeq:psihelthz}), (\ref{SUPeq:psihelphz}), and (\ref{SUPeq:alphamu2}), 
but also all azimuthal angles must be reflected before entering the matrix element formula:
%$\phi_{\pi}\to -\phi_{\pi}$,  
%$\phi_\mu\to -\phi_\mu$,
%$\phi_{\ZP}\to -\phi_{\ZP}$,  
%$\phi_\psi^{\ZP}\to -\phi_\psi^{\ZP}$,  
%$\phi_\mu^{\ZP}\to -\phi_\mu^{\ZP}$,
%$\alpha_p\to-\alpha_p$
%$\alpha_\mu\to -\alpha_\mu$,
$\phi_{\ZC}\to -\phi_{\ZC}$,  
$\phi_\psi^{\ZC}\to -\phi_\psi^{\ZC}$,  
$\phi_\mu^{\ZC}\to -\phi_\mu^{\ZC}$,
$\alpha_p^{\ZC}\to-\alpha_p^{\ZC}$
and
$\alpha^{\ZC}_\mu\to -\alpha^{\ZC}_\mu$
 \cite{Chilikin:2013tch}.

%It is clear from Eq.~(\ref{SUPPeq:total_matrixelement}) that various $\LambdaStarn$ 
%and $\ZP$ resonances
%interfere in the differential distributions.
%By integrating the matrix element squared over the entire phase space 
%the interferences cancel in the integrated rates unless
%the resonances belong to the same decay chain 
%and have the same quantum numbers.\footnote{For $\LambdaStarn-\ZP$, 
%the $\lambda_{\Lb}=+1/2$ interference terms have the opposite effect to  
%the $\lambda_{\Lb}=-1/2$ interference terms.}

\clearpage

% Author List ----------------------------                                                                                                                                                                                                                                                                                                
%  You need to get a new author list!                                                                                                                                                                                                                                                                                                    

%\input{LHCb_HD_authorlist_2014-06-20}
%\newpage
%\input{justification}
\newpage
\centerline{\large\bf LHCb collaboration}
\begin{flushleft}
\small
R.~Aaij$^{39}$,
C.~Abell{\'a}n~Beteta$^{41}$,
B.~Adeva$^{38}$,
M.~Adinolfi$^{47}$,
Z.~Ajaltouni$^{5}$,
S.~Akar$^{6}$,
J.~Albrecht$^{10}$,
F.~Alessio$^{39}$,
M.~Alexander$^{52}$,
S.~Ali$^{42}$
% ...

{\footnotesize \it
% ...
$ ^{5}$Clermont Universit{\'e}, Universit{\'e} Blaise Pascal, CNRS/IN2P3, LPC, Clermont-Ferrand, France\\
$ ^{6}$CPPM, Aix-Marseille Universit{\'e}, CNRS/IN2P3, Marseille, France\\
% ...
$ ^{10}$Fakult{\"a}t Physik, Technische Universit{\"a}t Dortmund, Dortmund, Germany\\
% ...
$ ^{38}$Universidad de Santiago de Compostela, Santiago de Compostela, Spain\\
$ ^{39}$European Organization for Nuclear Research (CERN), Geneva, Switzerland\\
% ...
$ ^{41}$Physik-Institut, Universit{\"a}t Z{\"u}rich, Z{\"u}rich, Switzerland\\
$ ^{42}$Nikhef National Institute for Subatomic Physics, Amsterdam, The Netherlands\\
% ...
$ ^{47}$H.H. Wills Physics Laboratory, University of Bristol, Bristol, United Kingdom\\
% ...
$ ^{52}$School of Physics and Astronomy, University of Glasgow, Glasgow, United Kingdom\\
% ...
}
\end{flushleft}
\end{document}